\documentclass[longauth]{aa}

\usepackage[english]{babel}
\usepackage{units}
\usepackage{graphicx}
\usepackage{amsmath}
\usepackage{amssymb}
\usepackage{caption}
\usepackage{subcaption}

\usepackage{tikz}
\usepackage{pgf}
\usepackage{3dplot} 

\usepackage{natbib}
\bibpunct{(}{)}{;}{a}{}{,} 

\definecolor{dark-gray}{gray}{0.3}

\newcommand{\matr}[1]{\mathbf{#1}}
\newcommand{\vect}[1]{\vec{\mathbf{#1}}}
\newcommand{\uvect}[1]{\mathbf{\hat{#1}}}

\newcommand{\figscale}{0.5}

\usepackage{subfig}

\usetikzlibrary{shapes,arrows}

\tikzstyle{block} = [rectangle, draw, fill=blue!20,
    text width=5em, text centered, rounded corners, minimum height=4em]
\tikzstyle{line} = [draw, -latex']
\tikzstyle{cloud} = [draw, rectangle,fill=red!20, node distance=3cm,
    minimum height=2em]

\begin{document}

\title{Detecting cosmic rays with the LOFAR radio telescope}

\date{Received 16 September 2013 / Accepted 4 November 2013}

\author{
P.~Schellart\inst{\ref{nijmegen}} \and
A.~Nelles\inst{\ref{nijmegen}\and \ref{nikhef}} \and
S.~Buitink\inst{\ref{kvi}\and\ref{nijmegen}}\and
A.~Corstanje\inst{\ref{nijmegen}}\and
J.~E.~Enriquez\inst{\ref{nijmegen}}\and
H.~Falcke\inst{\ref{nijmegen}\and\ref{nikhef}\and\ref{astron}\and\ref{mpifr}}\and
W.~Frieswijk\inst{\ref{astron}}\and
J.~R.~ H\"orandel\inst{\ref{nijmegen}\and\ref{nikhef}}\and
A.~Horneffer\inst{\ref{mpifr}}\and
C.~W.~James\inst{\ref{ecap}}\and
M.~Krause\inst{\ref{nijmegen}}\and
M.~Mevius\inst{\ref{astron}\and \ref{kapteyn}}\and
O.~Scholten\inst{\ref{kvi}}\and
S.~ter Veen\inst{\ref{nijmegen}}\and
S.~Thoudam\inst{\ref{nijmegen}}\and
M.~van den Akker\inst{\ref{nijmegen}}\and
A.~Alexov\inst{\ref{stsci}\and \ref{uva}}\and
J.~Anderson\inst{\ref{mpifr}}\and
I.~M.~Avruch\inst{\ref{sron}}\and
L.~B\"ahren\inst{\ref{nijmegen}}\and
R.~Beck\inst{\ref{mpifr}}\and
M.~E.~Bell\inst{\ref{caastro}}\and
P.~Bennema\inst{\ref{astron}}\and
M.~J.~Bentum\inst{\ref{astron}}\and
G.~Bernardi\inst{\ref{crat}\and \ref{skasa}}\and
P.~Best\inst{\ref{roe}}\and
J.~Bregman\inst{\ref{astron}}\and
F.~Breitling\inst{\ref{aip}}\and
M.~Brentjens\inst{\ref{astron}}\and
J.~Broderick\inst{\ref{soton}}\and
M.~Br\"uggen\inst{\ref{hamburg}}\and
B.~Ciardi\inst{\ref{mpifa}}\and
A.~Coolen\inst{\ref{astron}}\and
F.~de Gasperin\inst{\ref{hamburg}}\and
E.~de Geus\inst{\ref{astron}}\and
A.~de Jong\inst{\ref{astron}}\and
M.~de Vos\inst{\ref{astron}}\and
S.~Duscha\inst{\ref{astron}}\and
J.~Eisl\"offel\inst{\ref{tls}}\and
R.~A.~Fallows\inst{\ref{astron}}\and
C.~Ferrari\inst{\ref{nice}}\and
M.~A.~Garrett\inst{\ref{astron}\and \ref{leiden}}\and
J.~Grie\ss{}meier\inst{\ref{cnrs}}\and
T.~Grit\inst{\ref{astron}}\and
J.~P.~Hamaker\inst{\ref{astron}}\and
T.~E.~Hassall\inst{\ref{soton}\and \ref{jod}}\and
G.~Heald\inst{\ref{astron}}\and
J.~W.~T.~Hessels\inst{\ref{astron}\and \ref{uva}}\and
M.~Hoeft\inst{\ref{tls}}\and
H.~A.~Holties\inst{\ref{astron}}\and
M.~Iacobelli\inst{\ref{leiden}}\and
E.~Juette\inst{\ref{raiub}}\and
A.~Karastergiou\inst{\ref{ox}}\and
W.~Klijn\inst{\ref{astron}}\and
J.~Kohler\inst{\ref{mpifr}}\and
V.~I.~Kondratiev\inst{\ref{astron}\and \ref{lebedev}}\and
M.~Kramer\inst{\ref{mpifr}\and \ref{jod}}\and
M.~Kuniyoshi\inst{\ref{mpifr}}\and
G.~Kuper\inst{\ref{astron}}\and
P.~Maat\inst{\ref{astron}}\and
G.~Macario\inst{\ref{nice}}\and
G.~Mann\inst{\ref{aip}}\and
S.~Markoff\inst{\ref{uva}}\and
D.~McKay-Bukowski\inst{\ref{sodankyla}\and \ref{stfc}}\and
J.~P.~McKean\inst{\ref{astron}}\and
J.~C.~A.~Miller-Jones\inst{\ref{curtin}\and \ref{uva}}\and
J.~D.~Mol\inst{\ref{astron}}\and
D.~D.~Mulcahy\inst{\ref{mpifr}}\and
H.~Munk\inst{\ref{astron}}\and
R.~Nijboer\inst{\ref{astron}}\and
M.~J.~Norden\inst{\ref{astron}}\and
E.~Orru\inst{\ref{astron}}\and
R.~Overeem\inst{\ref{astron}}\and
H.~Paas\inst{\ref{groningen}}\and
M.~Pandey-Pommier\inst{\ref{lyon}}\and
R.~Pizzo\inst{\ref{astron}}\and
A.~G.~Polatidis\inst{\ref{astron}}\and
A.~Renting\inst{\ref{astron}}\and
J.~W.~Romein\inst{\ref{astron}}\and
H.~R\"ottgering\inst{\ref{leiden}}\and
A.~Schoenmakers\inst{\ref{astron}}\and
D.~Schwarz\inst{\ref{bielefeld}}\and
J.~Sluman\inst{\ref{astron}}\and
O.~Smirnov\inst{\ref{crat}\and \ref{skasa}}\and
C.~Sobey\inst{\ref{mpifr}}\and
B.~W.~Stappers\inst{\ref{jod}}\and
M.~Steinmetz\inst{\ref{aip}}\and
J.~Swinbank\inst{\ref{uva}}\and
Y.~Tang\inst{\ref{astron}}\and
C.~Tasse\inst{\ref{meudon}}\and
C.~Toribio\inst{\ref{astron}}\and
J.~van Leeuwen\inst{\ref{astron}\and \ref{uva}}\and
R.~van Nieuwpoort\inst{\ref{escience}\and \ref{astron}}\and
R.~J.~van Weeren\inst{\ref{cfa}}\and
N.~Vermaas\inst{\ref{astron}}\and
R.~Vermeulen\inst{\ref{astron}}\and
C.~Vocks\inst{\ref{aip}}\and
C.~Vogt\inst{\ref{astron}}\and
R.~A.~M.~J.~Wijers\inst{\ref{uva}}\and
S.~J.~Wijnholds\inst{\ref{astron}}\and
M.~W.~Wise\inst{\ref{astron}\and \ref{uva}}\and
O.~Wucknitz\inst{\ref{ubonn}\and \ref{mpifr}}\and
S.~Yatawatta\inst{\ref{astron}}\and
P.~Zarka\inst{\ref{meudon}}\and
A.~Zensus\inst{\ref{mpifr}}
}

\institute{
Department of Astrophysics/IMAPP, Radboud University Nijmegen, P.O. Box 9010, 6500 GL Nijmegen, The Netherlands\label{nijmegen}\and
Nikhef, Science Park Amsterdam, 1098 XG Amsterdam, The Netherlands\label{nikhef}\and
Netherlands Institute for Radio Astronomy (ASTRON), Postbus 2, 7990 AA Dwingeloo, The Netherlands\label{astron}\and
Max-Planck-Institut f\"{u}r Radioastronomie, Auf dem H\"ugel 69, 53121 Bonn, Germany\label{mpifr}\and
KVI, University Groningen, 9747 AA Groningen, The Netherlands\label{kvi}\and
ECAP, University of Erlangen-Nuremberg, 91058 Erlangen, Germany\label{ecap}\and
Astronomical Institute 'Anton Pannekoek', University of Amsterdam, Postbus 94249, 1090 GE Amsterdam, The Netherlands\label{uva}\and
Kapteyn Astronomical Institute, PO Box 800, 9700 AV Groningen, The Netherlands\label{kapteyn}\and
Leiden Observatory, Leiden University, PO Box 9513, 2300 RA Leiden, The Netherlands\label{leiden}\and
Jodrell Bank Center for Astrophysics, School of Physics and Astronomy, The University of Manchester, Manchester M13 9PL,UK\label{jod}\and
Astrophysics, University of Oxford, Denys Wilkinson Building, Keble Road, Oxford OX1 3RH\label{ox}\and
School of Physics and Astronomy, University of Southampton, Southampton, SO17 1BJ, UK\label{soton}\and
Max Planck Institute for Astrophysics, Karl Schwarzschild Str. 1, 85741 Garching, Germany\label{mpifa}\and
International Centre for Radio Astronomy Research - Curtin University, GPO Box U1987, Perth, WA 6845, Australia\label{curtin}\and
STFC Rutherford Appleton Laboratory,  Harwell Science and Innovation Campus,  Didcot  OX11 0QX, UK\label{stfc}\and
Institute for Astronomy, University of Edinburgh, Royal Observatory of Edinburgh, Blackford Hill, Edinburgh EH9 3HJ, UK\label{roe}\and
LESIA, UMR CNRS 8109, Observatoire de Paris, 92195   Meudon, France\label{meudon}\and
Argelander-Institut f\"{u}r Astronomie, University of Bonn, Auf dem H\"{u}gel 71, 53121, Bonn, Germany\label{ubonn}\and
Leibniz-Institut f\"{u}r Astrophysik Potsdam (AIP), An der Sternwarte 16, 14482 Potsdam, Germany\label{aip}\and
Th\"{u}ringer Landessternwarte, Sternwarte 5, D-07778 Tautenburg, Germany\label{tls}\and
Astronomisches Institut der Ruhr-Universit\"{a}t Bochum, Universitaetsstrasse 150, 44780 Bochum, Germany\label{raiub}\and
Laboratoire de Physique et Chimie de l' Environnement et de l' Espace, LPC2E UMR 7328 CNRS, 45071 Orl\'{e}ans Cedex 02, France\label{cnrs}\and
SRON Netherlands Insitute for Space Research, PO Box 800, 9700 AV Groningen, The Netherlands\label{sron}\and
Center for Information Technology (CIT), University of Groningen, The Netherlands\label{groningen}\and
Centre de Recherche Astrophysique de Lyon, Observatoire de Lyon, 9 av Charles Andr\'{e}, 69561 Saint Genis Laval Cedex, France\label{lyon}\and
ARC Centre of Excellence for All-sky astrophysics (CAASTRO), Sydney Institute of Astronomy, University of Sydney Australia\label{caastro}\and
University of Hamburg, Gojenbergsweg 112, 21029 Hamburg, Germany\label{hamburg}\and
Astro Space Center of the Lebedev Physical Institute, Profsoyuznaya str. 84/32, Moscow 117997, Russia\label{lebedev}\and
Centre for Radio Astronomy Techniques \& Technologies (RATT), Department of Physics and Electronics, Rhodes University, PO Box 94, Grahamstown 6140, South Africa\label{crat}\and
SKA South Africa, 3rd Floor, The Park, Park Road, Pinelands, 7405, South Africa\label{skasa}\and
Harvard-Smithsonian Center for Astrophysics, 60 Garden Street, Cambridge, MA 02138, USA\label{cfa}\and
Laboratoire Lagrange, UMR7293, Universit\`{e} de Nice Sophia-Antipolis, CNRS, Observatoire de la C\'{o}te d'Azur, 06300 Nice, France\label{nice}\and
Space Telescope Science Institute, 3700 San Martin Drive, Baltimore, MD 21218, USA\label{stsci}\and
Sodankyl\"{a} Geophysical Observatory, University of Oulu, T\"{a}htel\"{a}ntie 62, 99600 Sodankyl\"{a}, Finland\label{sodankyla}\and
Netherlands eScience Center, Science Park 140, 1098 XG, Amsterdam, The Netherlands\label{escience}\and
Fakult\"{a}t fur Physik, Universit\"{a}t Bielefeld, Postfach 100131, D-33501, Bielefeld, Germany\label{bielefeld}
}

\abstract{
The low frequency array (LOFAR), is the first radio telescope designed with the capability to measure radio emission from cosmic-ray induced air showers in parallel with interferometric observations. In the first $\sim \unit[2]{years}$ of observing, 405 cosmic-ray events in the energy range of $\unit[10^{16} - 10^{18}]{eV}$ have been detected in the band from $\unit[30 - 80]{MHz}$. Each of these air showers is registered with up to $\sim1000$ independent antennas resulting in measurements of the radio emission with unprecedented detail. This article describes the dataset, as well as the analysis pipeline, and serves as a reference for future papers based on these data. All steps necessary to achieve a full reconstruction of the electric field at every antenna position are explained, including removal of radio frequency interference, correcting for the antenna response and identification of the pulsed signal.
}

\maketitle

\section{Introduction}
\label{sec:introduction}
With the development of ever faster electronics and the increase in computational power, the construction of radio telescopes as large interferometric arrays of rather simple antennas opens a new window for observations. The low frequency array \citep[LOFAR;][]{LOFAR}, is the first large-scale implementation of this technique. In addition to producing the first high quality images at these low frequencies of $\unit[10-240]{MHz}$, LOFAR was designed to study short, pulsed signals in the time-domain. With a vast array of antennas observing the whole sky simultaneously, observations are not limited to a predefined direction, therefore providing optimal conditions for cosmic-ray detection.

Cosmic rays, accelerated charged particles from astrophysical sources, can be observed over several decades of energy. When cosmic rays of high energies reach the Earth, they do not reach the surface as primary particles, but instead interact with atmospheric nuclei. Thereby, a cascade of particles is created, consisting mostly of photons and a significant fraction of charged particles. While propagating through the atmosphere, the charged particles of this extensive air shower emit electromagnetic radiation, which adds up coherently for wavelengths comparable to the dimensions of the shower front \citep{RadioTheory}.

Already in the 1960s it was proven that cosmic ray-induced air showers emit nanosecond duration pulses with significant power in the MHz radio frequency range \citep{Allan1966,Jelley1965}, but due to lack of sufficiently sophisticated and fast electronics the technique was not pursued further. Only in the past decade interest in the detection technique was rekindled and successfully applied \citep{Falcke2008ICRC}. The proof of principle and large progress in the understanding of the emission was made at the LOFAR Prototype Experimental Station \citep[LOPES;][]{Falcke2005, Lopes2012} and further refined by measurements at the CODALEMA experiment \citep{Codalema2005}.

Similar to optical measurements of the fluorescence emission from atoms excited by interaction with the air shower, radio emission directly traces the longitudinal shower development, which is closely related to the type of the primary particle. Unlike optical fluorescence measurements, radio emission measurements are less dependent on observing conditions and can operate day and night matching the duty cycle of particle detector measurements.

Due to the very steep energy spectrum, measuring the highest-energy cosmic rays requires vast detector areas. Cost constraints therefore limit the density of detectors within this area giving a wide spacing between the individual antennas. Theoretical models describing the different emission mechanisms at play point to a very detailed and non-symmetrical emission pattern at ground level \citep{EVA, ZHAires, Selfas, CoREAS}. Testing these models therefore requires dense sampling of the electric field over a sufficiently large area.

LOFAR offers a high number of antennas clustered on an irregular grid, with increasingly large spacing between antenna clusters further away from the center. In the core of the array about 2300 antennas are installed within about $\unit[4]{km^{2}}$, which allows air showers to be measured with unprecedented spatial resolution. These measurements will contribute significantly to conclusively confirm theoretical models for the radio emission on a shower by shower basis, a goal previously unattainable due to lack of sufficiently high quality data.

Measurements and converging theoretical predictions of the expected radio signal from a cosmic-ray induced air shower give a short, nanosecond time-scale bi-polar pulse, which is mostly linearly polarized. This article describes the detection set-up and automated processing pipeline used at LOFAR to measure and identify these signals.

Starting with a description of the instrumental set-up at LOFAR in Sect.~\ref{sec:instrumental_setup}, an overview of the data reduction pipeline is given in Sect.~\ref{sec:pipeline}. Finally, Sect.~\ref{sec:dataset} describes the characteristics of the dataset obtained between June 2011 and April 2013.

The LOFAR dataset will be used in forthcoming publications to verify existing models for radio emission from air showers and to develop new techniques that use radio emission to measure important characteristics of the incoming particle, such as energy and mass.

\section{LOFAR}
\label{sec:instrumental_setup}

\begin{figure}
	\centering
	\includegraphics[width=\figscale\textwidth]{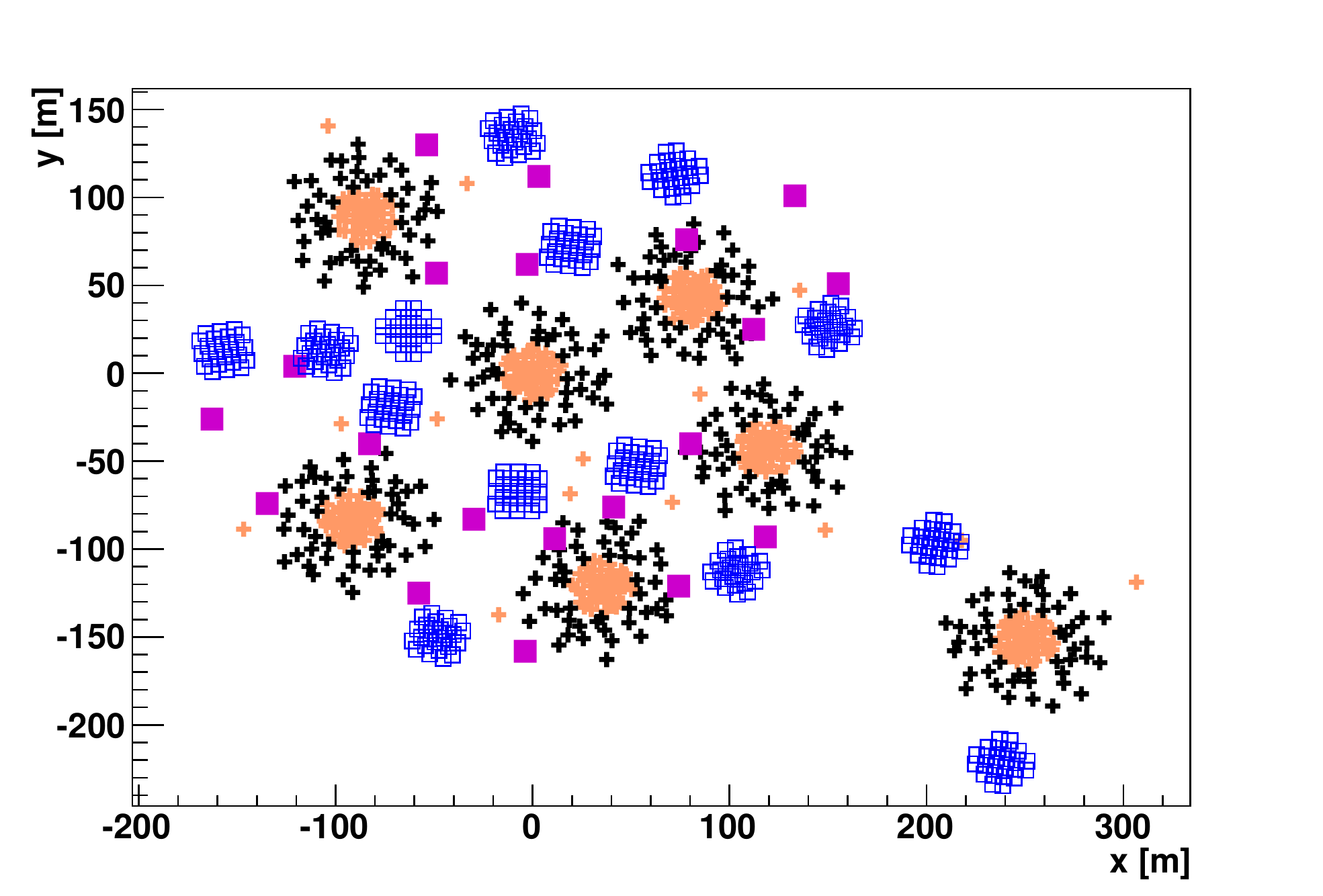}
	\caption[Superterp]{Layout of the center of LOFAR. The six stations to the left form the Superterp. The crosses indicate the LBA inner and outer antenna sets, respectively. The open squares show the positions of the HBA tiles, which are split into two groups per station. The filled squares indicate the positions of the LORA particle detectors.}
\label{fig:Superterp}
\end{figure}

LOFAR is a distributed radio telescope. Its antennas are distributed over northern Europe with the densest concentration in the north of the Netherlands, in the Province of Drenthe. The observation support center and processing facilities are also located near this central core. The antennas of LOFAR are grouped into \emph{stations}, each station taking the role of a single dish in a traditional radio interferometer array. A station consist of a number of low-band antennas (LBAs, $\unit[10-90]{MHz}$) and high-band antennas (HBAs, $\unit[110-240]{MHz}$). The 24 stations within the $\sim\unit[2]{km}$ wide core are distributed in an irregular pattern that maximizes $uv$-coverage, or spatial frequencies for standard interferometric observations. The 16 additional Dutch remote stations are distributed with increasing distance to the core.  International stations are currently located in Germany, France, the United Kingdom, and Sweden, giving LOFAR a maximum baseline of $\unit[1292]{km}$ for interferometric observations. Core stations and remote stations consist of 96 LBAs plus 48 HBAs. International stations have 96 LBAs and 96 HBAs. At the center of the LOFAR core six stations are located in a roughly $\unit[320]{m}$ diameter area, called the \emph{Superterp}, providing both the shortest baselines for interferometric observations  and the densest population of antennas ideal for cosmic-ray observations. While every LOFAR station is equipped with the necessary electronics to observe cosmic rays, the current data set is taken with the central 24 stations, where additional information from particle detectors is available (see Sect.~\ref{sec:lora}). The positions of the antennas of the seven most central LOFAR stations are shown in Fig.~\ref{fig:Superterp}.
\subsection{The antennas}
\begin{figure}
	\centering
	\includegraphics[width=\figscale\textwidth]{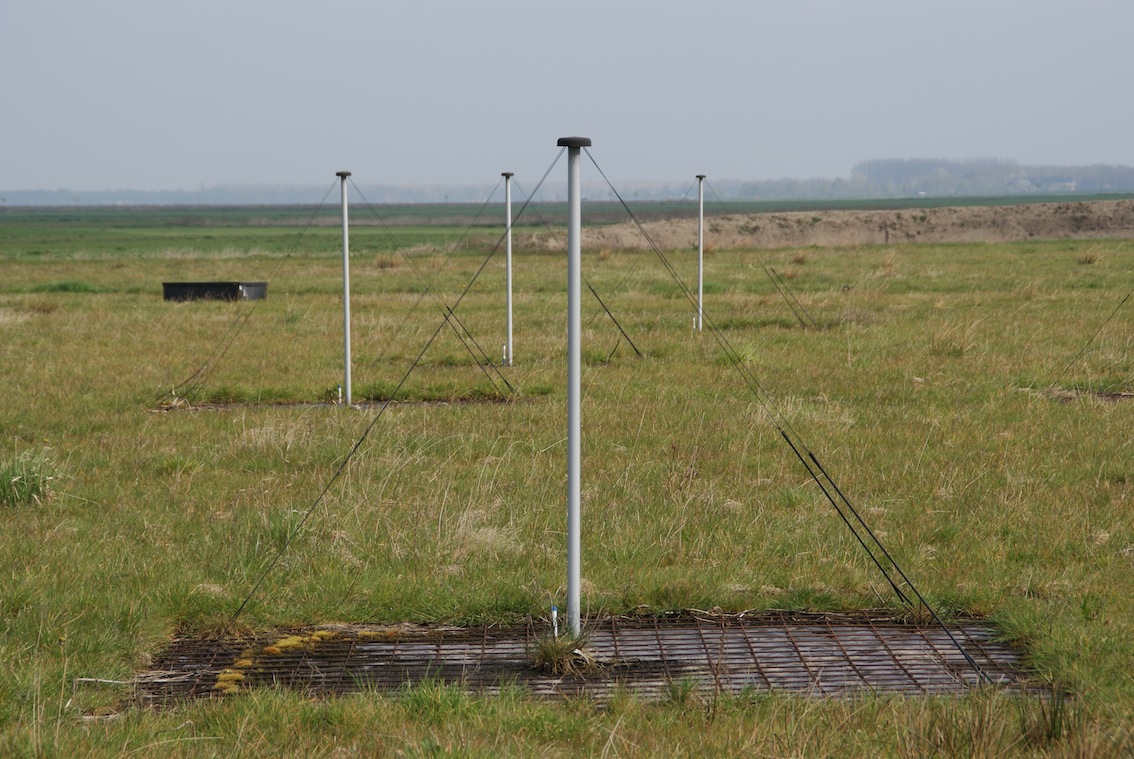}
	\caption[LBA]{Low-band antennas at the central core of LOFAR, the Superterp. In the background the black box of a LORA particle detector can be seen.}
\label{fig:LORA}
\end{figure}

\begin{figure}
\centering
\begin{tikzpicture}[scale=0.8]
\path (0, 0) -- (-3, 0) node[left]{W};
\draw[dashed, thick] (-3, 0) -- (3, 0) node[right]{E};
\path (0, 0) -- (0, -3) node[below]{S};
\draw[dashed, thick] (0, -3) -- (0, 3) node[above]{N};
\draw[thick, ->] [rotate=135] (0, -3) -- (0, 3) node[left]{$X$};
\draw[thick, ->] [rotate=225] (0, -3) -- (0, 3) node[right]{$Y$};
\draw[thick, ->] (0, 0) -- (1.5, 0) node[above]{$\hat{\mathbf{e}}_{x}$};
\draw[thick, ->] (0, 0) -- (0, 1.5) node[right]{$\hat{\mathbf{e}}_{y}$};
\draw (0.5, 0) arc (0:45:0.5);
\path (0.5, 0) arc (0:35:0.5) node[right]{$45^{\circ}$};
\end{tikzpicture}
\caption{Geometry of the LBA. The X and Y dipoles are oriented NE-SW and NW-SE respectively. This is rotated by 225 degrees with respect to the standard local Cartesian coordinate system used in Sect.~\ref{sec:coordinate_transformation}.}
\label{fig:instrumental_polarization_coordinates}
\end{figure}
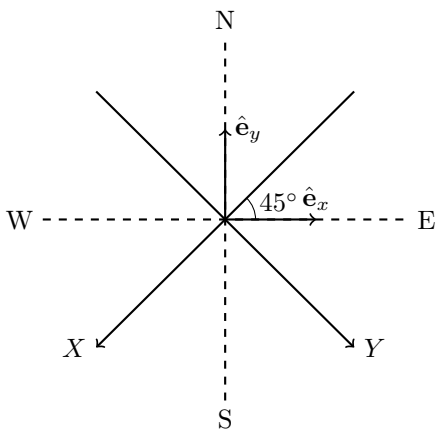

The LBAs are the main tool for cosmic-ray detection. An LBA consists of two orthogonal inverted V-shaped dipoles, each with a length of $\unit[1.38]{m}$. These are supported by a central polyvinyl chloride pole, which holds the low-noise amplifier and guides the signal cables, as shown in Fig.~\ref{fig:LORA}. The dipoles $X$ and $Y$, that make up each antenna, are oriented southwest to northeast (SW-NE) and southeast to northwest (SE-NW), as can be seen in Fig.~\ref{fig:instrumental_polarization_coordinates}.

The low-noise amplifier has an intentional impedance mismatch with the antenna. This mismatch, combined with the characteristic length of the dipoles, makes the system sensitive in a broad band from $\unit[10-90]{MHz}$. In principle, this allows observations from the ionospheric cutoff up to the start of the commercial FM radio band. For most observations the frequency range is limited by a combination of selectable hardware and software filters to $\unit[30-80]{MHz}$ to suppress strong Radio Frequency Interference (RFI) in the outer bands. The LBAs are designed to be sky noise limited after RFI has been removed \citep{LBADesign}. After amplification the signals from the individual dipoles are transmitted through coaxial cables to the electronics cabinet located at every station.

The HBAs have been optimized for a frequency band of $\unit[110-240]{MHz}$. The design clusters 16 antenna elements into a ``tile'', the signals from these elements are amplified and combined in an analog beam-former. This means that while the LBAs are sensitive to the whole sky the HBAs are most sensitive within the $\sim 20^{\circ}$ of the tile-beam, of which the direction is chosen at the start of every observation. This results in a smaller effective area for cosmic-ray observations, as the measurement will only be optimal if the direction of the cosmic ray happens to coincide with the beam direction of the observation. Therefore, the analysis of HBA data and their interesting higher frequency range requires a different approach for cosmic-ray studies. Results of these measurements will be described in a later publication.

\subsection{The transient buffer boards}
After being forwarded to the electronics cabinet the signals of the LBAs are again amplified, filtered, and digitized by a $\unit[12]{bit}$ A/D converter with a sampling frequency of $\unit[200]{MHz}$\footnote{A 160 MHz clock is also available.}. Due to signal path limitations in the Dutch stations only 48 dual-polarized or 96 single-polarized antennas can be processed at a given time. For the dual-polarized option the antennas are grouped into an inner and an outer set, which has to be chosen before an observation.

For astronomical observations the data are then beam-formed and sent to the central processing facility. In addition, there is the possibility to store a snapshot of the original data. Every station is equipped with ring-buffers, the so called Transient Buffer Boards (TBBs). These continuously store the last $\unit[1.3]{s}$ of raw data (an extension to $\unit[5]{s}$ is currently being deployed). When triggered, the contents of the TBBs are frozen, read out via the Wide Area Network and stored on disk for further analysis. The trigger can be generated based on various parameters in an FPGA\footnote{Field Programmable Gate Array.} at the local receiver unit. Alternatively, the trigger can be generated by an array of particle detectors (see Sect.~\ref{sec:lora}) or received from outside of LOFAR. Currently, the main trigger for cosmic-ray observation is provided by the particle detectors. Later, a radio self-trigger will be implemented, using the current dataset as a training set to deduce trigger criteria, so that the FPGA trigger can be run independently at every LOFAR station. These criteria have to reduce false triggers to limit the data rate. Using every LOFAR station individually will dramatically increase the effective area.

Essential for measuring cosmic rays with LOFAR as a radio telescope is that the whole process of triggering and storing radio-pulse data can take place without interfering with the ongoing observations.

\begin{figure}
	\centering
	\includegraphics[width=\figscale\textwidth]{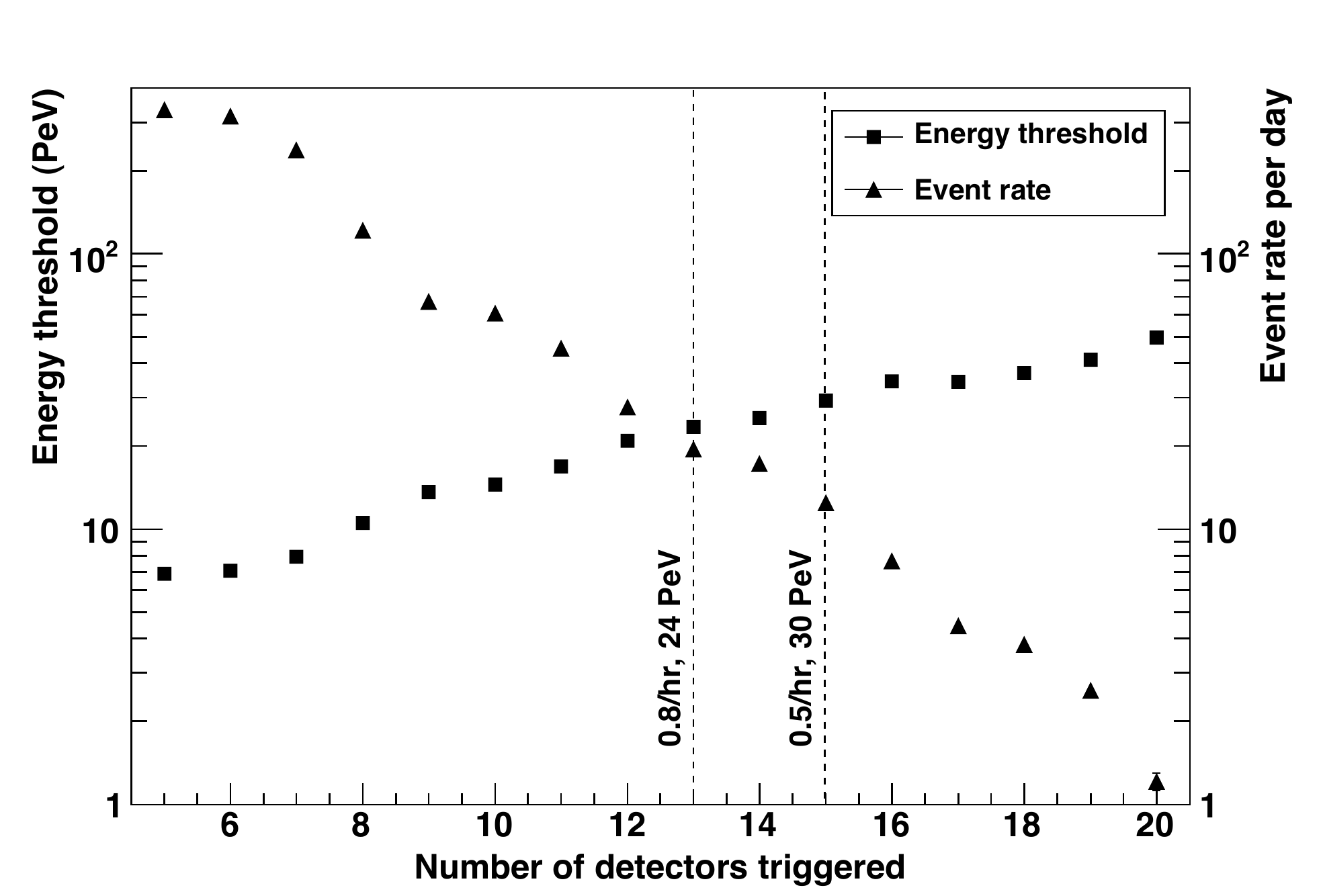}
	\caption[LORA sensitivity]{Energy threshold in PeV (left) and the event rate per day (right) are shown as a function of the number of triggered particle detectors. Two possible trigger conditions are indicated with the dotted lines.}
	\label{fig:lora_sens}
\end{figure}

\subsection{The LOFAR Radboud Air Shower Array}
\label{sec:lora}
LORA, the LOFAR Radboud Air Shower Array, is an array of particle detectors co-located with the center of LOFAR. The array provides a reconstruction of basic parameters of recorded air showers, such as the direction and the position of impact, as well as the energy of the incoming cosmic ray \citep{Thoudam2013}. It also provides the time of arrival, which is used to trigger the read-out of the radio antennas.

LORA consists of 20 detector units distributed on the Superterp, as shown in Fig.~\ref{fig:Superterp}. Each detector contains two scintillators ($\unit[0.45]{m^{2}}$, type: NE 114), which are individually read out through a photomultiplier tube. The detectors are inside weatherproof shelters and have been tested to not create any interference at radio frequencies.

Conditions at which triggers are sent to LOFAR can be adjusted to match the desired energy threshold. There are two constraints on the desired rate: the rate of events interesting for radio observations has to be maximized, while the network load on the LOFAR system has to be kept low in order to avoid interfering with the primary observation. A trigger in a single detector is generated when a particle signal of more than $4\sigma$ above the noise is registered. In order to only detect air showers a coincidence of several detectors is needed. Events of less than $\unit[10^{16}]{eV}$ have a very low probability to be observable in radio above the sky-noise level. The energy threshold and the corresponding event rate are shown in Fig.~\ref{fig:lora_sens} as the function of the number of triggered detectors. Requiring triggers in 13 detectors yields a threshold energy of $\unit[2.4\cdot10^{16}]{eV}$, with an average trigger-rate of $\unit[0.8]{events/hour}$. This trigger rate has been selected as the optimal setting for the observations.

\subsection{Observations}
\label{sec:observations}
After the commissioning phase LOFAR is to be used on a proposal-based schedule. Proposals are open to the community for imaging or beam-formed observations, as well as TBB observations. Some fraction of the observing time is reserved for participating consortia and key science projects. The LOFAR cosmic ray key science project (CRKSP) is one of six LOFAR key science projects.

To maximize the duty cycle TBB observations can be run in the background of all other observations that do not need the full network bandwidth. This does however mean that the array configuration is determined by the primary observation, therefore the amount of data in a specific array configuration (such as the selection of LBA or HBA antenna type) available for analysis is not determined by the cosmic-ray project itself, except when LOFAR is otherwise idle and the observing configuration can be chosen freely.

During the observation, triggers from LORA are received by the LOFAR control system. The system checks whether a dump from the TBBs is allowed. If so, the ring-buffers are frozen and a specified block of data around the trigger time is dumped to disk. For each cosmic-ray event $\unit[2.1]{ms}$ of radio data are stored, which corresponds to $\unit[77]{MB}$ per station. This provides sufficient frequency resolution for high quality RFI cleaning while minimizing data transfer and storage requirements.

Every evening, the data-files are archived at LOFAR and compressed for transport. They are stored in the Long Term Archive \citep{LOFAR}, from where they can be retrieved for data analysis.

\section{Reconstruction of cosmic-ray data}
\label{sec:pipeline}
All newly recorded data are processed every evening, after having been copied via the network to the processing cluster of the Astrophysics department at the Radboud University Nijmegen. In addition to the \textsc{hdf5}\footnote{A tree-like file format \citep{Alexov2013}.} files, containing the data of one LOFAR station each, the recorded data from the particle detectors and a trigger log file are transferred. With this information an automated pipeline is run. The pipeline is based on the task oriented PyCRTools framework consisting of fast low-level C++ routines embedded in Python for maximum flexibility. All results are stored in a PostgreSQL database for subsequent data mining analysis. The goal of the processing pipeline is to autonomously identify a full set of physics quantities for each air shower detected with LOFAR. The pipeline is optimized to identify those nanosecond pulses that are not generated by terrestrial sources.

All data are first processed per station, i.e.\ per file. The set of files received for a single trigger form an \emph{event}.  When the data from one station pass the criteria for containing a cosmic-ray signal (see Sect.~\ref{sec:pd}), the corresponding event is called a \emph{cosmic-ray event}. It is not necessary to observe a pulse in all stations, only the stations with a significant signal are used in a combined analysis.

\subsection{Pipeline structure}
\begin{figure}
\centering
\begin{tikzpicture}[node distance = 1.8cm, auto]
    \node [cloud] (file) {Data file};
    \node [block, below of=file] (lora_block) {Block selection};
    \node [cloud, right of=lora_block] (lora_timestamp) {LORA timestamp};
    \node [block, below of=lora_block] (phase_calibration) {Phase calibration};
    \node [cloud, right of=phase_calibration] (delay_tables) {LOFAR tables};
    \node [block, below of=phase_calibration] (rfi_cleaning) {RFI cleaning};
    \node [block, left of=rfi_cleaning, node distance=3cm] (rfi_identification) {RFI identification};
    \node [block, below of=rfi_cleaning] (gain_calibration) {Gain calibration (relative)};
    \node [cloud, right of=gain_calibration] (galactic_noise) {Galactic noise};
    \node [block, below of=gain_calibration] (beamforming) {Beam forming};
    \node [block, below of=beamforming] (unfolding) {Unfolding antenna response};
    \node [cloud, right of=unfolding] (antenna_model) {Antenna model};
    \node [block, below of=unfolding] (pulse_finding) {Pulse detection};
    \node [block, below of=pulse_finding] (direction_fit) {Direction fitting};
    \node [block, below of=direction_fit] (projecting) {Coordinate transformation};
    \node [block, left of=pulse_finding, node distance=3cm] (update_direction) {Update direction};
    \node [block, below of=projecting] (physics_parameters) {Extracting pulse parameters};
    \path [line] (file) -- (lora_block);
    \path [line] (lora_block) -- (phase_calibration);
    \path [line] (phase_calibration) -- (rfi_cleaning);
    \path [line] (rfi_cleaning) -- (gain_calibration);
    \path [line] (gain_calibration) -- (beamforming);
    \path [line] (beamforming) -- (unfolding);

    \path [line] (file) -| (rfi_identification);
    \path [line] (rfi_identification) -- (rfi_cleaning);

    \path [line] (unfolding) -- (pulse_finding);
    \path [line] (pulse_finding) -- (direction_fit);
    \path [line] (direction_fit) -- (projecting);

    \path [line] (direction_fit) -| (update_direction);
    \path [line] (update_direction) |- (unfolding);

    \path [line] (projecting) -- (physics_parameters);

    \path [line, dashed] (lora_timestamp) -- (lora_block);
    \path [line, dashed] (delay_tables) -- (phase_calibration);
    \path [line, dashed] (galactic_noise) -- (gain_calibration);
    \path [line, dashed] (antenna_model) -- (unfolding);
\end{tikzpicture}
\caption[Pipeline structure]{General structure of the analysis pipeline. Rectangles represent input and rounded squares are processing steps.}
\label{fig:pipeline_structure}
\end{figure}
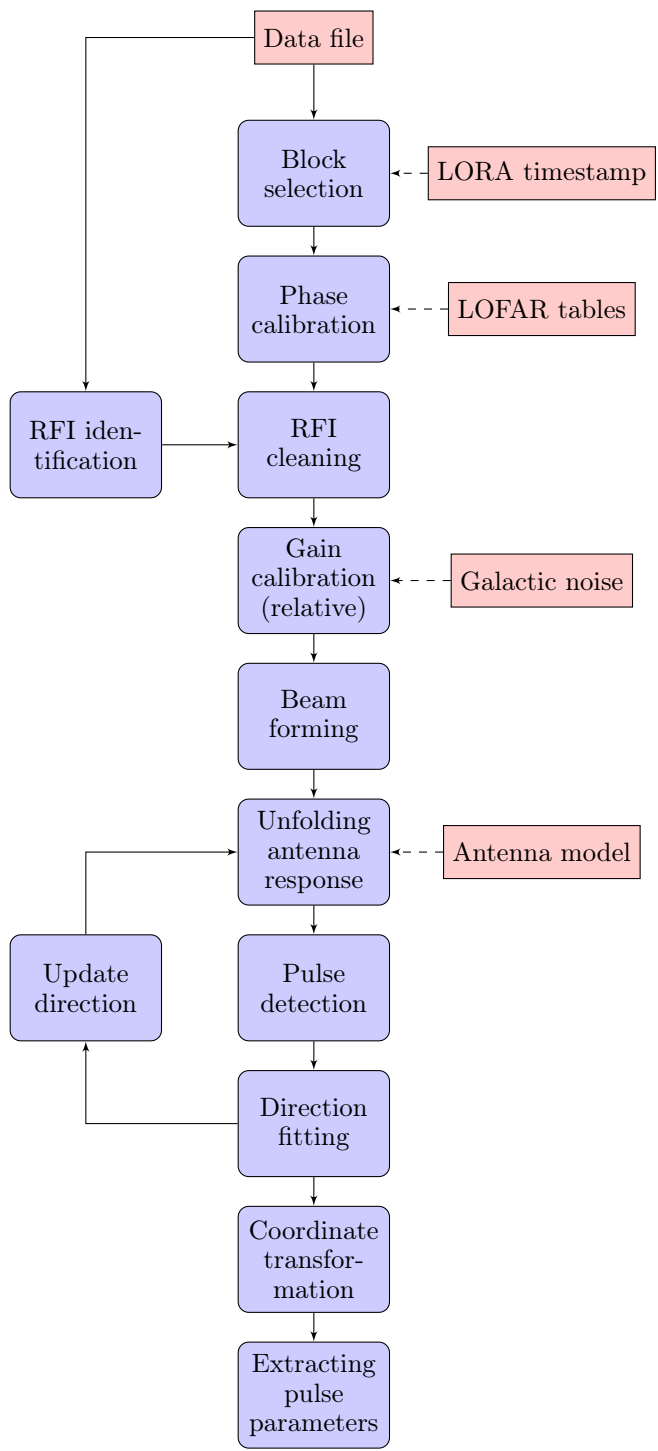

The reconstruction pipeline comprises a number of steps that will be individually explained in the following sections. An overview of the steps and the overall structure is depicted in Fig.~\ref{fig:pipeline_structure}.

\subsection{Preparing the data}
Before proceeding to extract the cosmic-ray signal from the data, some preparatory steps have to be performed. Knowledge about the system is applied in the form of calibration procedures, the data are cleaned of narrowband-transmitters, and antennas that show malfunctions are flagged.

\subsubsection{Timing offsets and phase calibration}
There are known signal path differences between the LOFAR antennas. Measured differences of cable lengths between the antennas are corrected for up to the $\unit[5]{ns}$ sample level already at the stations before the data are written to disk. Additionally, relative time offsets between the antennas are corrected for at sub-sample accuracy using standard LOFAR calibration tables. These tables are generated by phase-calibrating on the strongest astronomical radio sources and are regularly tested and updated if necessary \citep{LOFAR}. Sub-sample corrections are applied as phase offsets to the Fourier transformed signal in the cosmic-ray pipeline, before processing it in the data analysis.

\subsubsection{RFI cleaning}
\begin{figure}
\centering
\begin{subfigure}[b]{\figscale\textwidth}
	\includegraphics[width=0.9\textwidth]{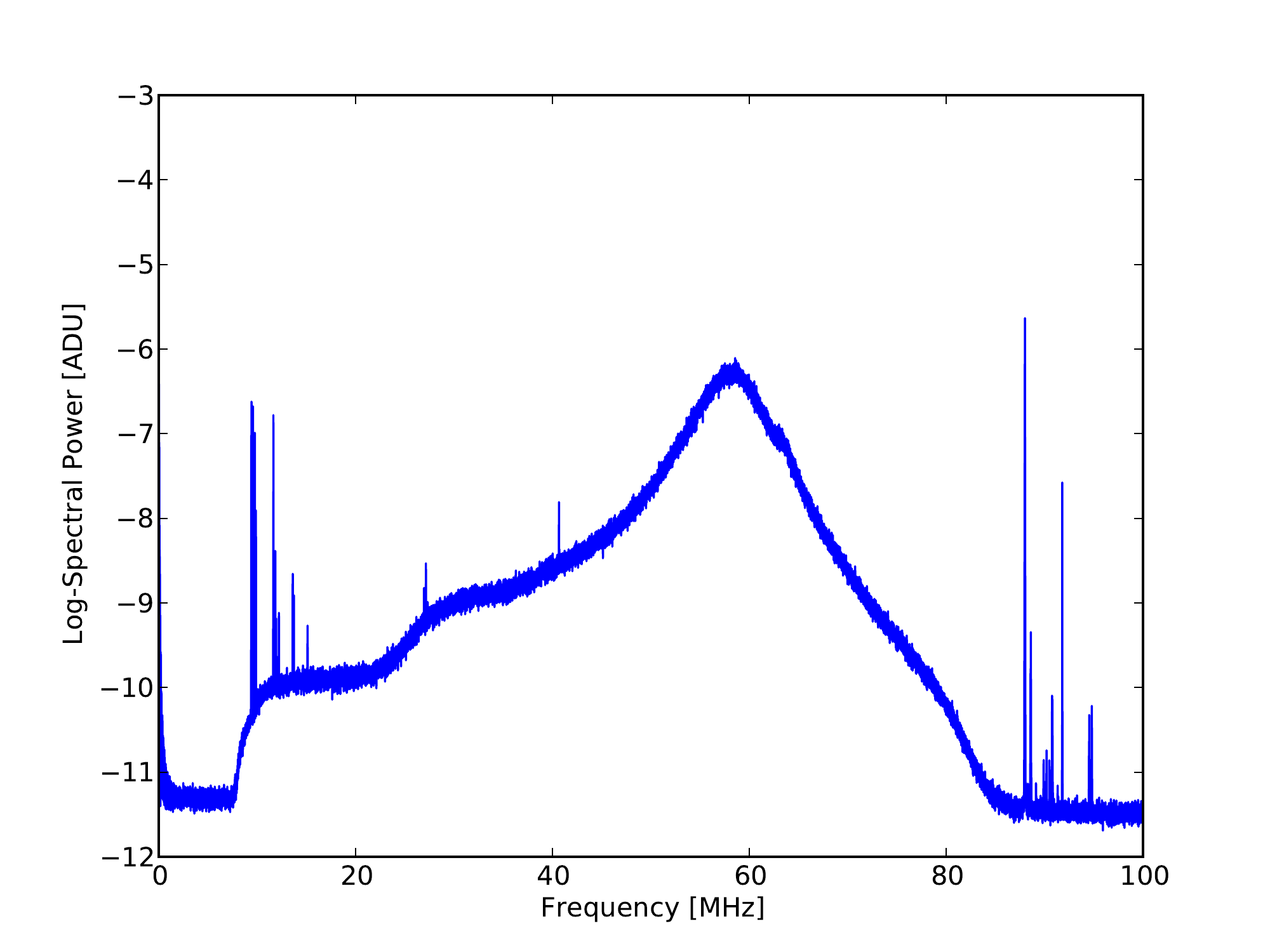}
	\label{fig:spectrum_org}
\end{subfigure}
\begin{subfigure}[b]{\figscale\textwidth}
\includegraphics[width=0.9\textwidth]{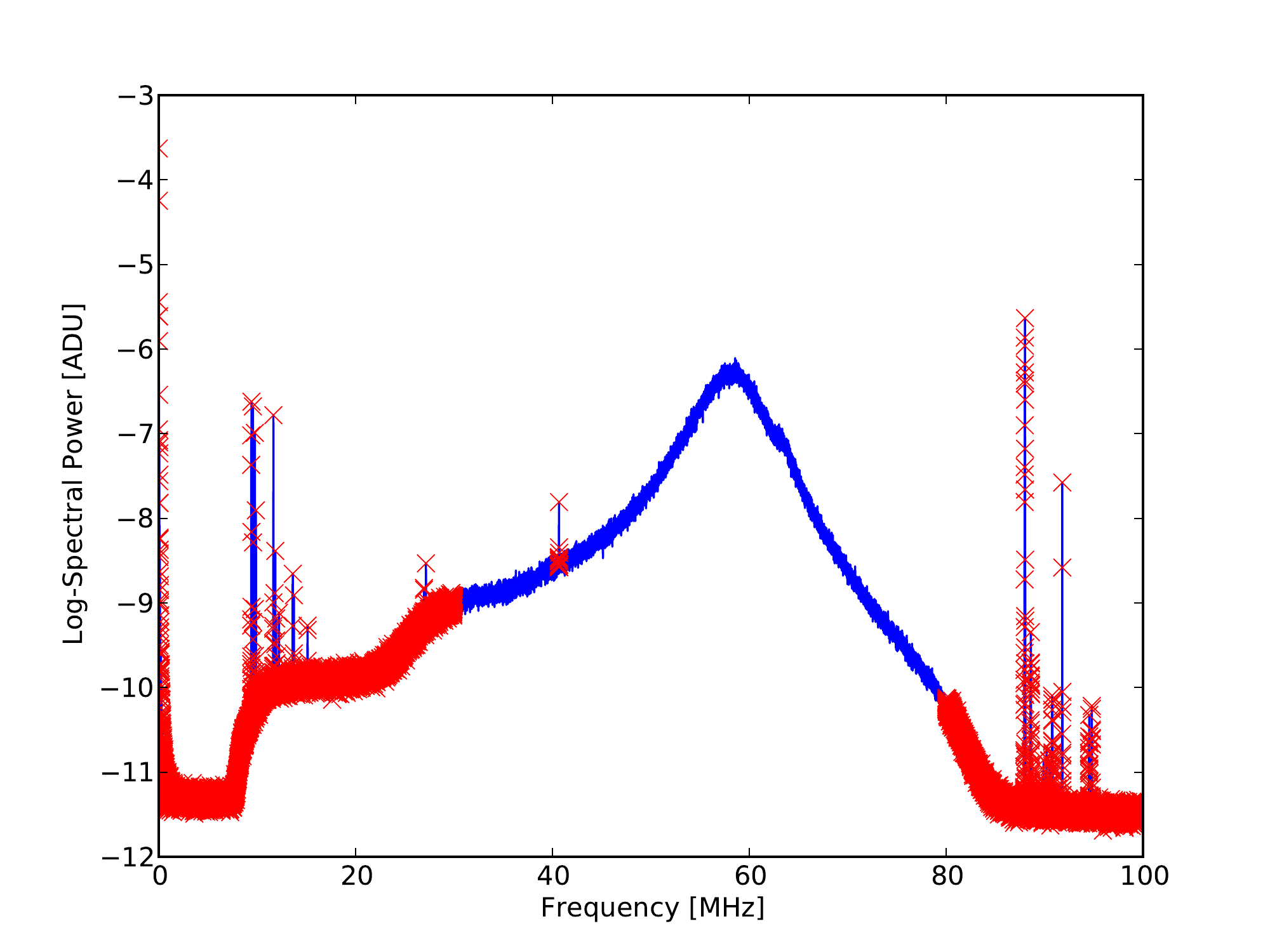}
\label{fig:spectrum_flag}
\end{subfigure}
\begin{subfigure}[b]{\figscale\textwidth}
\includegraphics[width=0.9\textwidth]{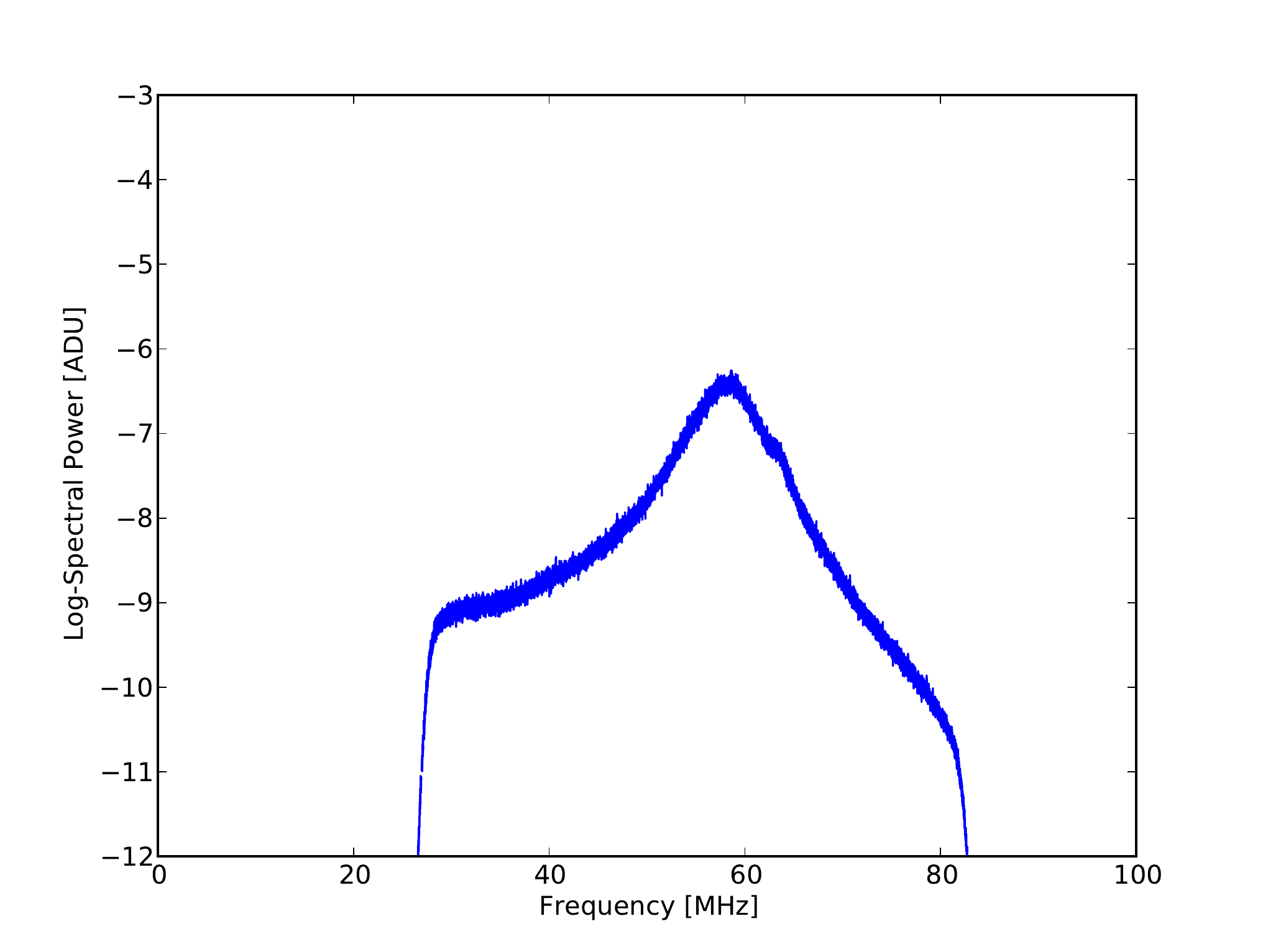}
\label{fig:spectrum_clean}
\end{subfigure}
\caption{The average spectrum of a typical LBA event. The raw data (\emph{top}), with flagged contaminated channels (\emph{middle}), cleaned and clipped to $\unit[30-80]{MHz}$ (\emph{bottom}).}
\label{fig:spectrum_lba}
\end{figure}

Narrow-band RFI in the time series signal can be revealed by making an average power spectrum. An example is shown in the top panel of Fig.~\ref{fig:spectrum_lba}, where most of the strong RFI is visible outside the $\unit[30-80]{MHz}$ range. The average power spectrum is created by averaging the square of the absolute value of the Fourier transform over several blocks of data. The block size can be freely chosen within the full data length to obtain a desired frequency resolution; here $2^{16}$ samples are used, giving a resolution of $\sim\unit[3]{kHz}$, enough to resolve most RFI lines. A reasonable data length is needed for this procedure to produce a stable average, which sets the limit for the chosen block length to be stored from the TBBs, as mentioned in Sect.~\ref{sec:observations}. In order to minimize artificial side lobes a half-Hann window is applied to the first and last 10\% of each trace prior to the Fourier transformation.

The standard approach to RFI cleaning (or RFI flagging) is to identify peaks sticking out significantly above the overall spectral shape, also called the \emph{baseline}, and set the corresponding Fourier component amplitudes to zero. However, this requires `a priori' knowledge of the baseline. While the baseline can be obtained through a smoothing or fitting procedure, this is often not stable in the presence of strong RFI, requiring an iterative approach.

An alternative approach to RFI cleaning uses the phase information in the complex-valued spectrum instead. If an RFI transmitter is measured in all antennas, the phase difference, or relative phase, between each pair of antennas will be a constant value as function of time with a small non constant random noise contribution. Note that the exact value of the constant, which only depends on the geometric delay between antennas, is not relevant, only its non time-varying nature. When no transmitter is present, the relative phase is expected to be both random and time varying, as the signal then consists of the added signals from many incoherent sources on the sky with additional random noise. Therefore, RFI can be identified by looking at the stability of phase differences between antennas over time.
For each antenna-dipole $j=0, 1, \ldots, 95$ in a station and data block $k$, the phase spectrum is calculated as
\begin{equation}
\phi_{j,k}(\omega) = \arg(x_{j,k}(\omega)),
\end{equation}
where $x_{j,k}(\omega)$ is the complex frequency component $\omega$ of the spectrum.

Subtracting the phase of one of the antennas as reference antenna gives the relative phases and results in a set of phases for every frequency channel, one for each block of data. Only one reference antenna is used and this is taken to be the one with median power to avoid selecting a broken antenna.

The average phase is defined as
\begin{equation}
\bar{\phi}_{j}(\omega) = \arg\left(\sum_{k=0}^{N-1} \exp(i\phi_{j,k}(\omega))\right),
\end{equation}
and the phase variance as
\begin{equation}
s_{j}(\omega) = 1 - \frac{1}{N}\left|\sum_{k=0}^{N-1} \exp(i\phi_{j,k}(\omega))\right|,
\end{equation}
where $N$ is the number of data blocks.

\begin{figure}
      \centering
      \includegraphics[width=\figscale\textwidth]{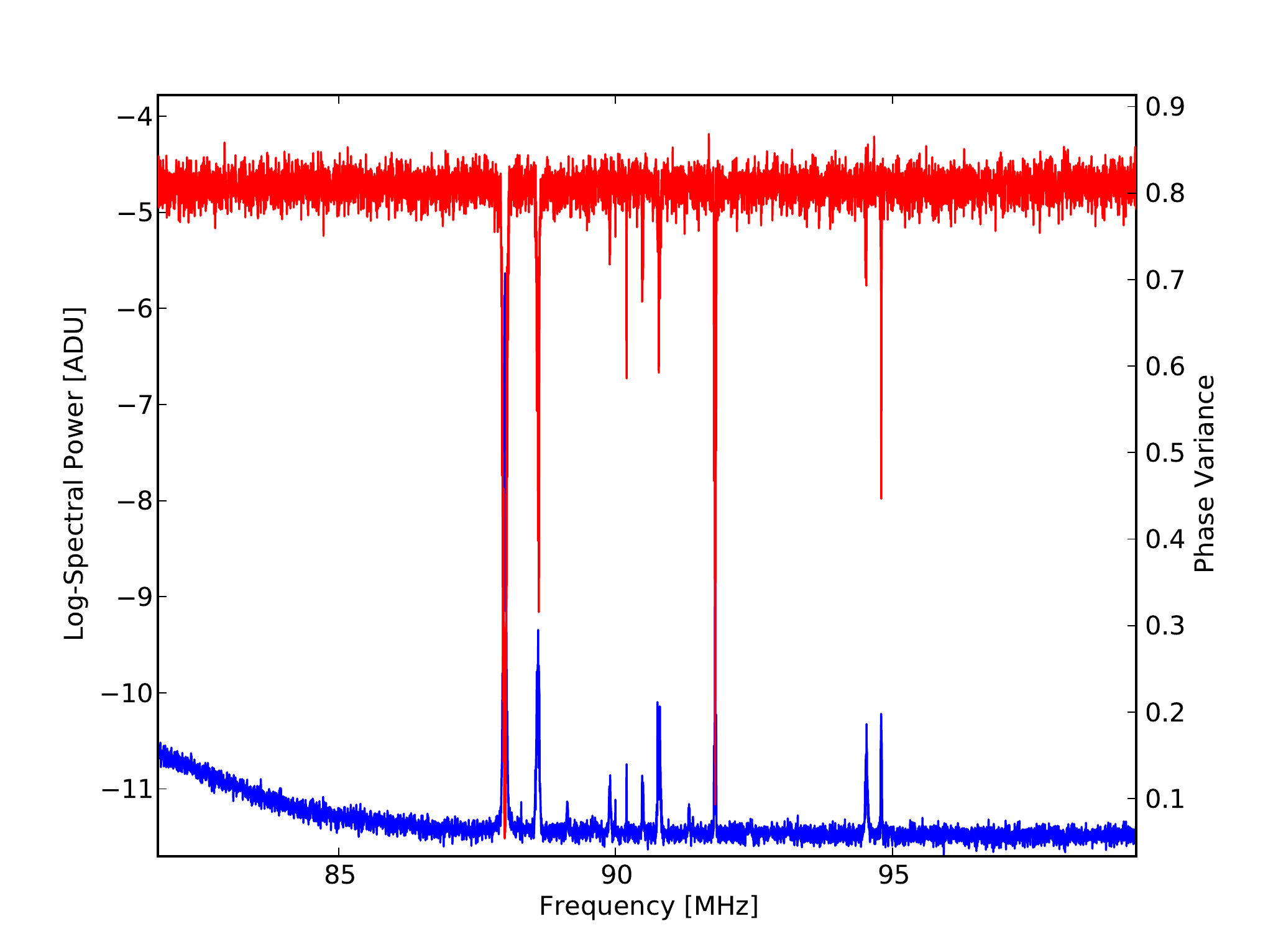}
      \caption[Phase variance]{Average LBA spectrum (bottom, left axis) with the corresponding phase variance (top, right axis). RFI lines can clearly be identified in the phase variance with peaks toward lower values, representing more stable phase differences between antennas over time.}
\label{fig:phase_variance}
\end{figure}

For completely random phases one expects $s_{j}(\omega)\approx1$ as opposed to $s_{j}(\omega)=0$ when all phases are equal. The phase variance per frequency channel will now either be at a value close to 1, including some random `noise', or at a significantly lower level. The latter reveals the presence of a radio transmitter, as shown in Fig.~\ref{fig:phase_variance}, where a contaminated part of the spectrum is shown with the corresponding phase variance.

Since RFI lines will result in peaks toward smaller values of the phase variance, and noise has no preferred peak direction, calculating the standard deviation $\sigma$ in this plot only for values above the median will ensure a stable result. All frequencies that have a phase variance of at least $5\sigma$ below the median\footnote{Assuming a Gaussian distribution, $\sigma$ can be estimated by sorting the data points, and comparing the value at 95 percentile to the median. This difference amounts to $\sim\unit[1.64]{\sigma}$.} are flagged as containing RFI. Additionally a $\unit[30-80]{MHz}$ bandpass filter is applied, flagging the most heavily RFI polluted low and high frequency parts of the bandwidth by default. To prevent pulse-ringing the $\unit[30-80]{MHz}$ block filter is first convolved with a, $\sigma_{\mathrm{tapering}} = \unit[2.5]{MHz}$, Gaussian\footnote{This effect also occurs when flagging large blocks of RFI but this does not happen in practice and so no tapering window is applied for this case.}. After removing the flagged channels, the resulting cleaned spectrum is shown in the bottom panel of Fig.~\ref{fig:spectrum_lba}.

\begin{figure}
	\centering
	\includegraphics[width=\figscale\textwidth]{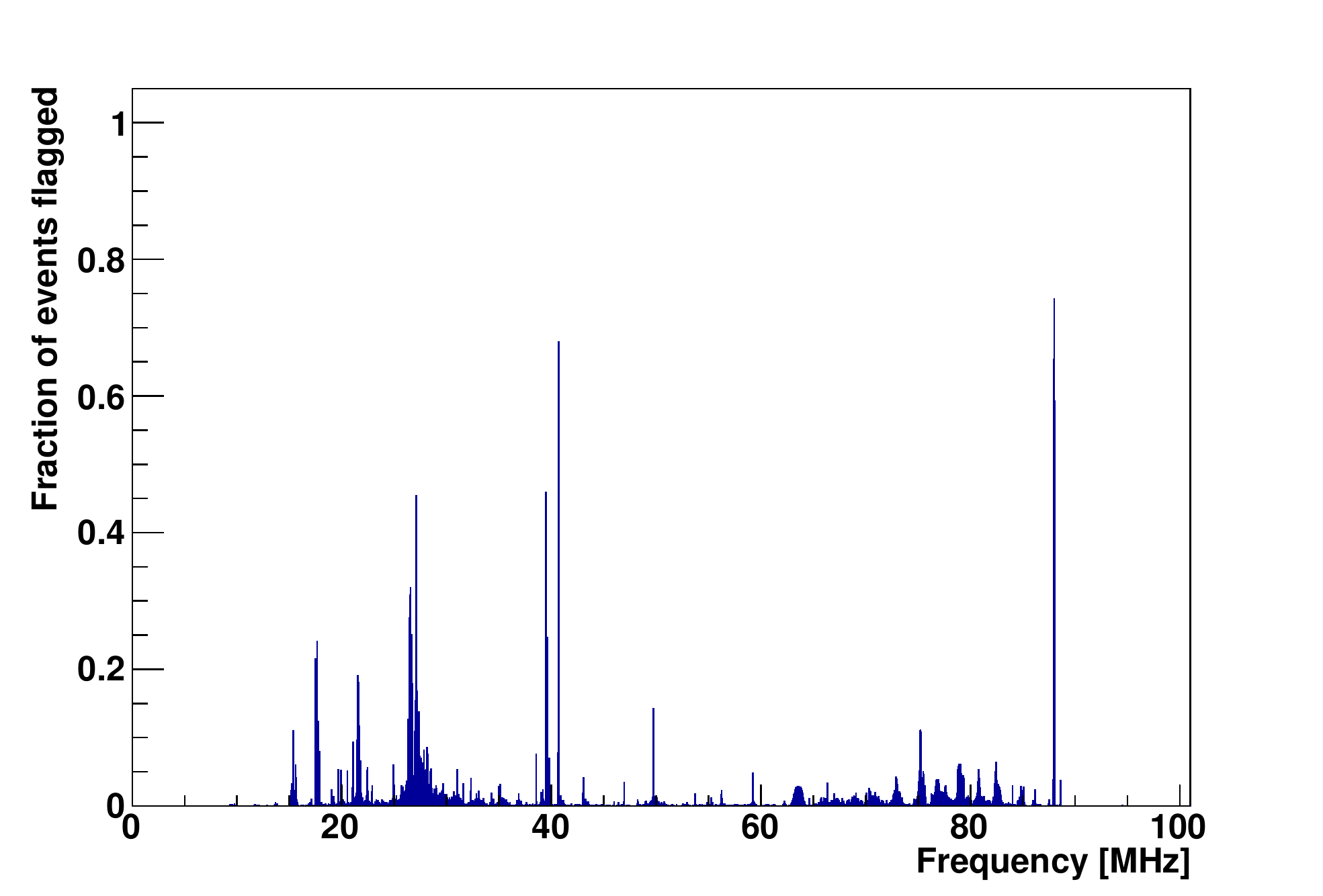}
	\caption[RFI statistics]{Fraction of events that is affected by narrow-band RFI in each of the $\sim\unit[3]{kHz}$ frequency channels as function of  frequency.}
	\label{fig:rfichannels}
\end{figure}

\begin{figure}
	\centering
	\includegraphics[width=\figscale\textwidth]{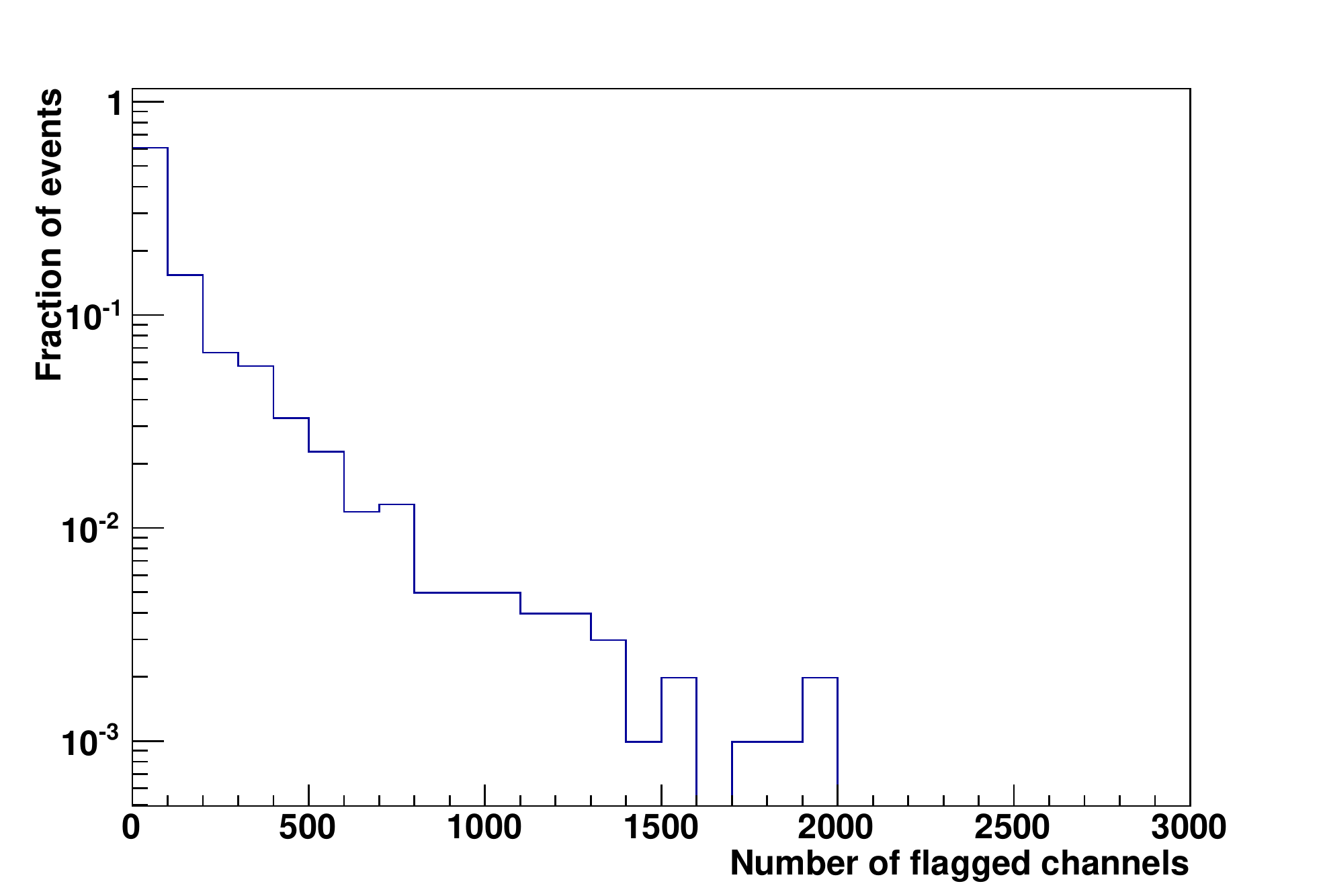}
	\caption[RFI statistics2]{Relative fraction of events with a certain number of flagged channels. Over 60\% of the events have less than 100 channels ($\approx\unit[300]{kHz})$ flagged out of the full used bandwidth of more than 16000 channels.}
	\label{fig:rfifrac}
\end{figure}

In general, there is very little RFI at the LOFAR Superterp. A lot of effort has been made to remove local sources that could disturb the LOFAR measurements and a protected zone has even been established \citep{Offringa2013}. This relative quietness is illustrated in Fig.~\ref{fig:rfichannels}. It shows the result of the RFI cleaning for all events for all frequencies. While every event has some RFI, no single RFI line is present in every event. Within the $\unit[30-80]{MHz}$ band, there are only two lines that are present in more than 40\% of the events. In total there are rarely events with more than 2\% flagged channels out of the more than 32000 frequency channels in a block of data. This is is shown in Fig.~\ref{fig:rfifrac}, where the total fraction of events is plotted against the number of flagged channels.

\subsubsection{Flagging bad antennas}
Occasionally, one or more antennas give invalid signals, e.g.\ due to hardware malfunction. To identify these \emph{bad antennas} the integrated spectral power is calculated
\begin{equation}
\label{eq:integrated_spectral_power}
P = \int_{\unit[30]{MHz}}^{\unit[80]{MHz}} |x(\omega)|^{2} d\omega,
\end{equation}
where $x(\omega)$ is the $\omega$ frequency component of the cleaned spectrum.
The power in every antenna is required to be in the range of one half to two times the median power from all antennas. Antennas outside this range are marked as bad and excluded from further analysis.

\subsubsection{Absolute gain calibration}
There are ongoing efforts for an absolute calibration of the voltage traces of LOFAR and therefore the reconstructed electric field. Those efforts will be described in a forthcoming publication and include calibration on astronomical sources, terrestrial transmitters, and already conducted dedicated measurement campaigns, similar to those performed at other experiments, e.g. \cite{Nehls2008}. Once implemented, the reconstruction pipeline will deliver calibrated electric field strengths and their polarization components for all events. However, significant progress in understanding the mechanisms of radio emission in air showers can already be made with a relative calibration.

\subsubsection{Relative gain calibration}
\begin{figure}
	\centering
	\begin{subfigure}[b]{\figscale\textwidth}
		\includegraphics[width=\textwidth]{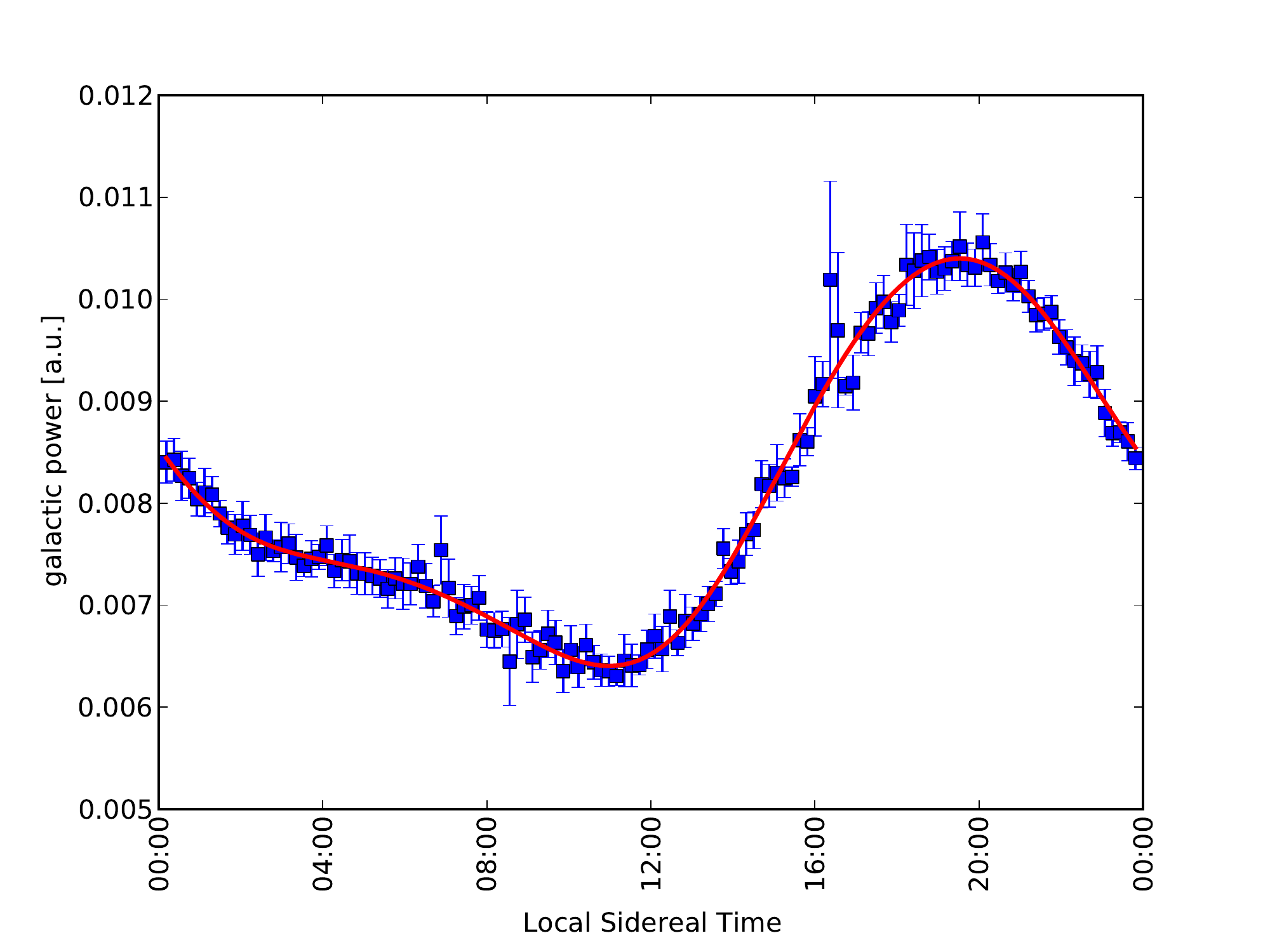}
	\end{subfigure}
	\begin{subfigure}[b]{\figscale\textwidth}
		\includegraphics[width=\textwidth]{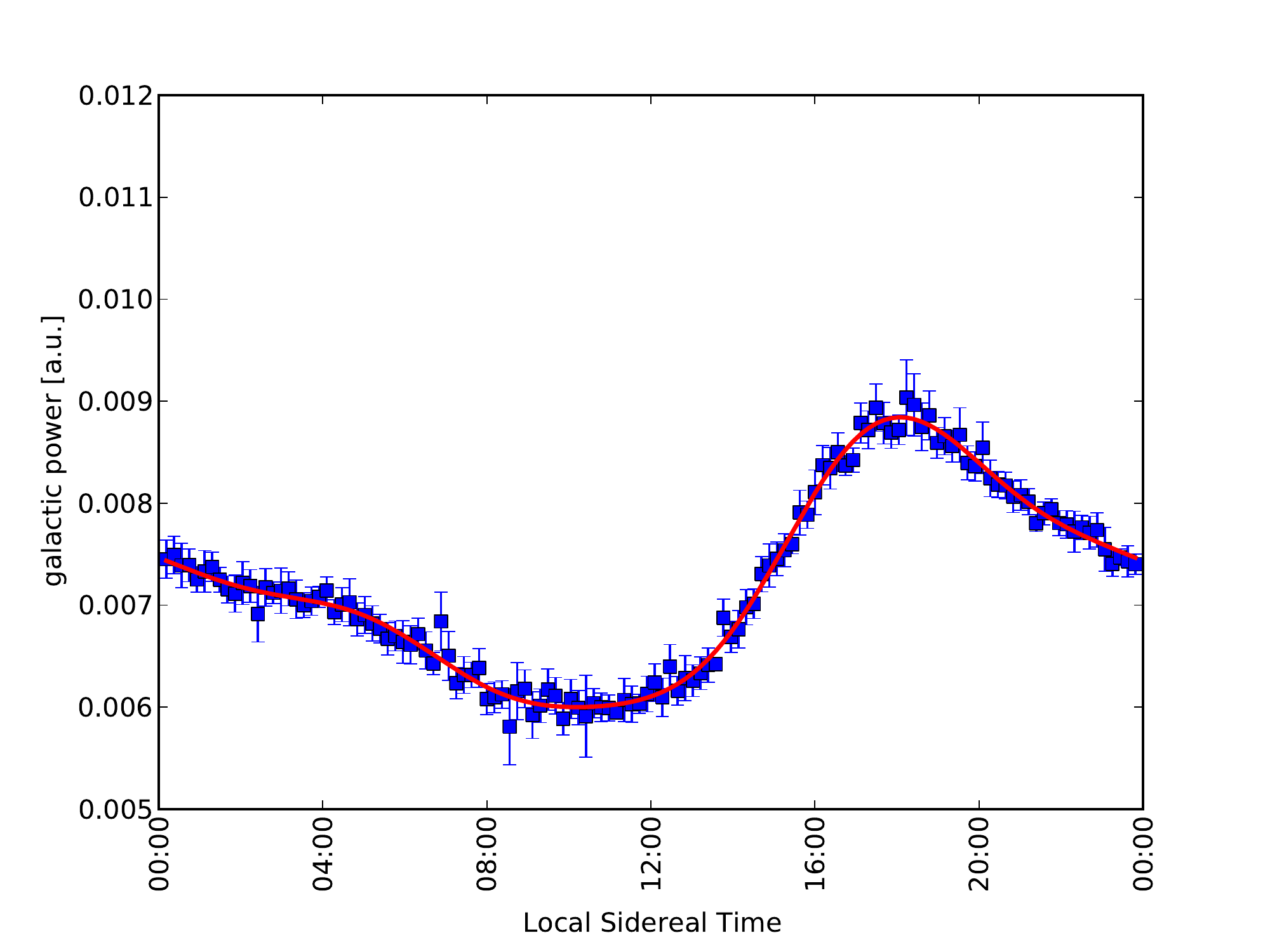}
	\end{subfigure}
	\caption[GalaxyFit1]{Integrated spectral power normalized to the bandwidth, after RFI cleaning, as a function of local sidereal time for the X (NE-SW) (\emph{top}) and Y (NW-SE) (\emph{bottom}) instrumental antenna polarizations. Also shown is the fitted second order Fourier transform (solid line). The uncertainties on the data still include systematic effects due to the set-up itself, as well as possible artifacts of the RFI cleaning, when having certain frequencies that are contaminated in a significant fraction of the data.}
\label{fig:galaxy}
\end{figure}

The LBA measurement is dominated by sky noise, which in turn is dominated by the Galaxy moving through the antenna beam pattern. Therefore, the noise as seen by each antenna is a function of the Local Sidereal Time (LST) and can be used to correct for differences in gain between antennas. Instead of correcting all antennas at all times to a fixed value, which would be over- or underestimating the noise at certain times, the received power can be normalized to a LST-dependent reference value. In Fig.~\ref{fig:galaxy} the integrated spectral power (equation \ref{eq:integrated_spectral_power}), after RFI cleaning, is given as a function of LST for the instrumental polarization $X$ and $Y$. The data have been retrieved from all cosmic-ray events measured within the first year of data-taking.
One can define a reference value for the integrated spectral power as a function of LST by fitting a function to these data points. Since the movement of the Galaxy through the antenna beam pattern is periodic by nature it is fitted with the 2nd order Fourier series
\begin{equation}
P_{\mathrm{ref}}(t) = \frac{a_{0}}{2} + \sum_{n=1}^{2} a_{n} \sin(nt) + b_{n} \cos(nt),
\end{equation}
thereby avoiding artificial jumps in the fit at 0:00 LST. The time $t$ is given in units of radian here.
This results in a gain correction for each antenna as
\begin{equation}
x'(\omega) = \sqrt{\frac{P_{\mathrm{ref}}(t)}{P(t)}} x(\omega),
\end{equation}
where the square root is needed, because the correction is applied to the amplitude spectrum.

\subsection{Identifying cosmic-ray signals}
\label{sec:pd}
After cleaning and calibration of the data, the central element of the pipeline is the identification and characterization of the radio pulse as the signal of the air shower.
\subsubsection{Using information from the particle detectors}
\begin{figure}
	\centering
	\includegraphics[width=\figscale\textwidth]{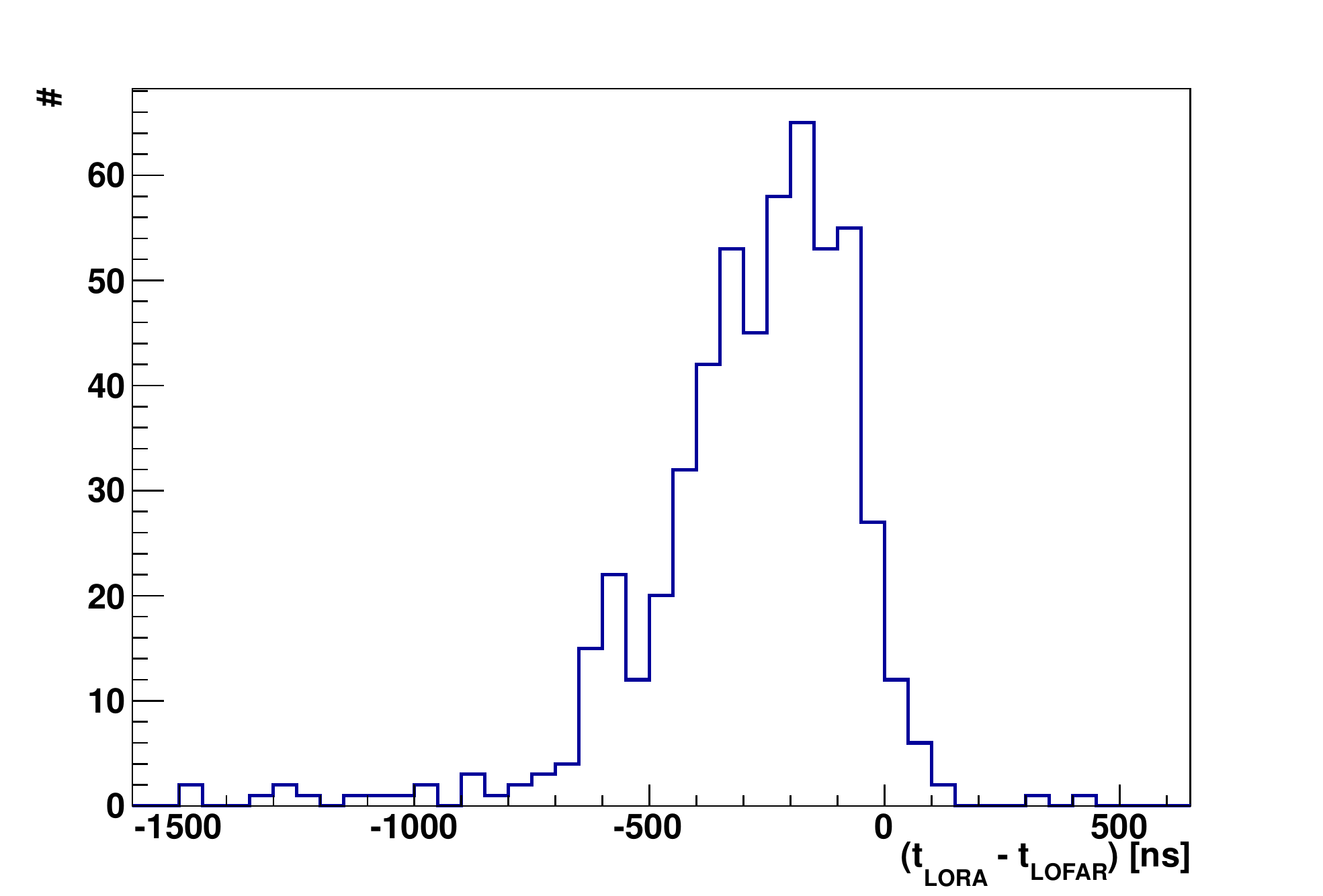}
	\caption[Pulse location with respect to LORA time]{Difference in time between the time of a pulse identified in the radio signal and the trigger time set by the signal in the particle detectors. This plot shows the distribution summed over all Superterp stations. }
\label{fig:lorawindow}
\end{figure}

In order to restrict the search for the radio pulse to a smaller region in the trace, the information from the trigger time of the particle detectors is used. Figure \ref{fig:lorawindow} shows the difference in time between the trigger from the particle detectors and the pulse location in the radio data obtained from a search with a large window. The distribution shows a clear peak at the region of the coincidences at an offset of  $\unit[253\pm168]{ns}$. In absolute timing the offset between LORA and LOFAR is $\unit[10253]{ns}$, of which $\unit[10000]{ns}$ are already accounted for in the triggering system.

Average offset is obtained by fitting a Gaussian to the distribution of pulse positions with respect to the trigger time. This is only an approximation, as the real offset per event depends on the position of the core and the incoming direction of the air shower. Also, effects due to the propagation of particles and radiation in the atmosphere can play a role. The overall difference is due to the fact that both detectors operate independently on different timing systems. Both are based on GPS timing, but correct for drifts ($<\unit[20]{ns}$) in different ways and have a differing absolute time. The spread on the differences is however sufficiently small for Superterp stations to not require additional synchronization of the two systems. Stations further away can have larger offsets due to the signal travel time, which can be corrected for after a reconstruction of the shower.

These measurements allow for the pulse search to be restricted to a small fraction of the full time trace, limiting the chance to pick up random noise fluctuations.

\subsubsection{Finding candidate events}
The trigger threshold of the scintillator array is chosen to be lower than the threshold to detect a radio signal.  This ensures a full sample, but also makes it necessary to identify in a first quality check whether there is a detectable signal present. Therefore, per antenna polarization, the signals are first \emph{beamformed} in the direction reconstructed from the data of the particle detectors.
This direction is given in the local Cartesian coordinate frame of the station by $\vect{n}$ and the position of each antenna $j$ is given by $\vect{r}_{j}$. A planar wavefront arriving at the phase center $(0, 0, 0)$ at time $t=0$ will arrive at antenna $j$ with a delay given by
\begin{equation}
\Delta t_{j} = - \frac{1}{c}\frac{\vect{n}\cdot\vect{r}_{j}}{|\vect{n}|} = - \frac{1}{c} \uvect{e}_{n} \cdot \vect{r}_{j},
\end{equation}
where $c$ is the speed of light. The beamformed signal, in frequency space, in this direction is then given by
\begin{equation}
x_{\mathrm{bf}}(\omega) = \sum_{j=0}^{N_a} x_{j}(\omega) e^{2\pi i \omega \Delta t_{j}},
\end{equation}
where $x_{j}(\omega)$ is frequency component $\omega$ of the Fourier transform of the signal from antenna $j$ and $N_a$ is the number of antennas. The inverse Fourier transform gives the beamformed time series signal. Due to beamforming any signal coming from the direction of the air shower is amplified by a factor $N_{a}$ in amplitude while uncorrelated noise is only amplified by a factor $\sqrt{N_{a}}$. Therefore, if no significant signal is detected in the beamformed trace, the event very unlikely contains a cosmic-ray signal strong enough to be detected at single dipole level by the rest of the pipeline. Thus, the analysis of the data of that station is aborted.

\begin{figure}
	\centering
	\includegraphics[width=\figscale\textwidth]{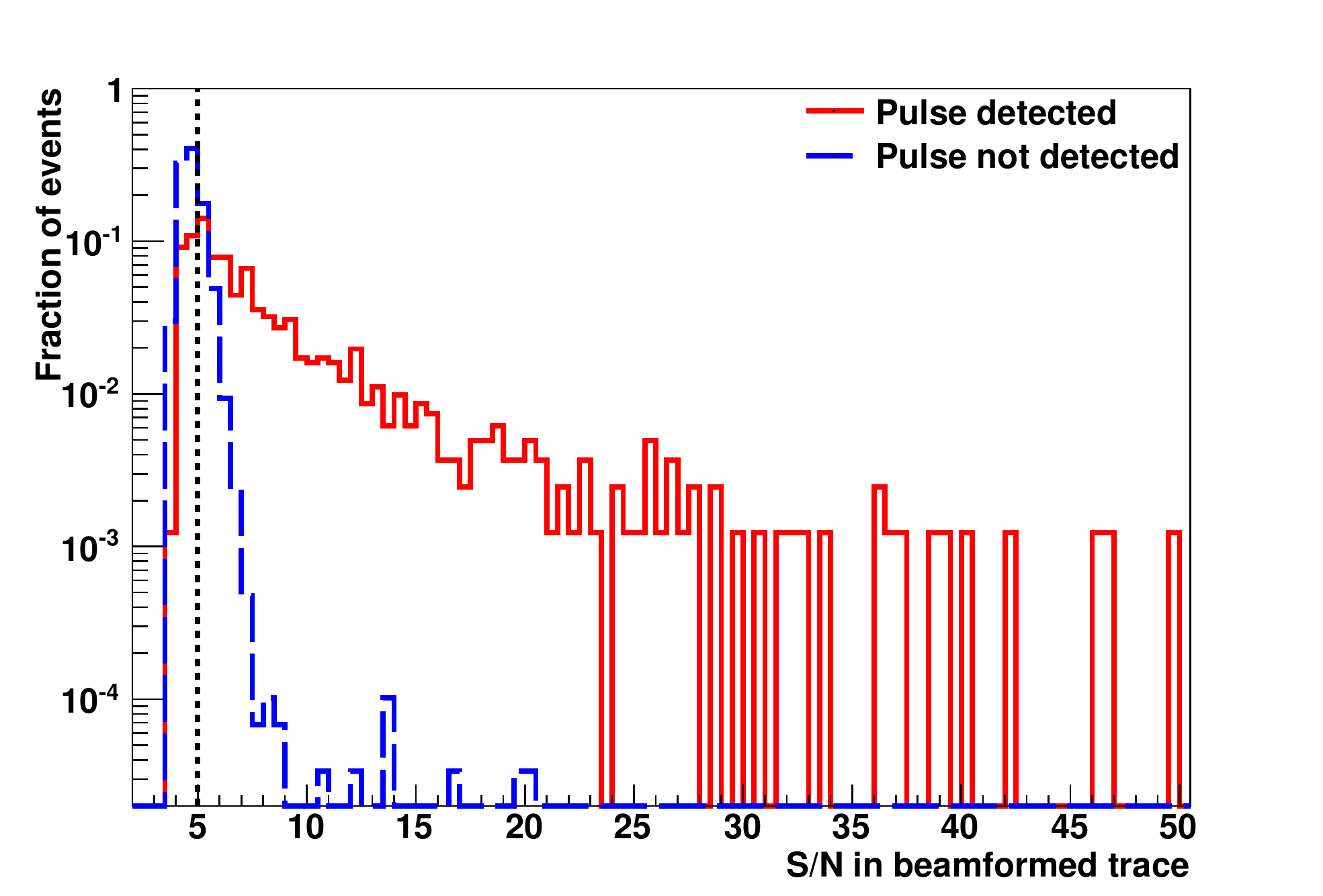}
	\caption[Beamformed pulse height]{Distribution of the signal-to-noise ratio (S/N) in initial beamforming. The S/N is defined the ratio of the peak amplitude of the beam-formed trace and the RMS of this trace. Two cases are separated:  a cosmic-ray event was ultimately detected by the pipeline (solid line) or not (dashed line). The initial cut, which is applied in the pipeline, is indicated by the dotted line. }
\label{fig:Beamformedpulseheight}
\end{figure}

To test this assumption, Fig.~\ref{fig:Beamformedpulseheight} shows the distribution of the peak amplitude in the beamformed signal per station, distinguishing between events in which ultimately a cosmic-ray was identified and those in which there was not. The peak amplitude is normalized by the root mean square of the trace, as a proxy for the noise contribution. From this it can be seen that the fraction of events where a strong signal is observed in the beamformed trace is significantly higher for stations where eventually a cosmic-ray signal is detected. All events in the tail of the non-detected distribution were visually inspected and identified as broad-band RFI, with pulses differing significantly in shape from those of cosmic rays and directions ultimately deviating significantly from the direction as measured with the particle detectors. This distribution shows that an initial filtering based on a moderate signal-to-noise of beamformed pulses is a quick and effective way to filter out those events that are potentially interesting, as well as further narrowing the search window per antenna reducing false positives for pulse detection.

\subsubsection{Correction for the antenna response}
The sensitivity of the LOFAR LBA is a complex function of both frequency and direction. Correcting for this \emph{antenna pattern}, i.e.\ unfolding, requires an initial guess for the pulse direction and in turn may influence the position of the pulse in time and thus the direction by changing the phase at which each frequency arrives. Therefore the correction has to be done in an iterative loop as indicated in Fig.~\ref{fig:pipeline_structure}. Each iteration starts with an increasingly accurate signal direction and proceeds by unfolding the antenna pattern, pulse detection, and direction fitting. The loop is concluded when the direction no longer significantly changes, which usually happens in less than $\sim 5$ iterations.

For the antenna pattern of the LBA a simulation is used, which is made using the software WIPL-D \citep{Kolundzija2011} and a customized software model of the electronics chain.

From the impedance and radiation pattern in a transmitting situation the open circuit voltage is calculated as a function of frequency and direction for an incoming plane wave with an electric field strength of $\unit[1]{V/m}$. The equivalent circuit of the antenna in a receiving situation is a voltage source with an internal resistance equal to the antenna impedance. This is combined with measured data of the amplifier directly behind the antenna. The result of the model is the output voltage of the amplifier over a $\unit[75]{\Omega}$ resistor\footnote{Matched to the impedance of the coaxial cables connecting the antenna to the station electronics cabinet.}.

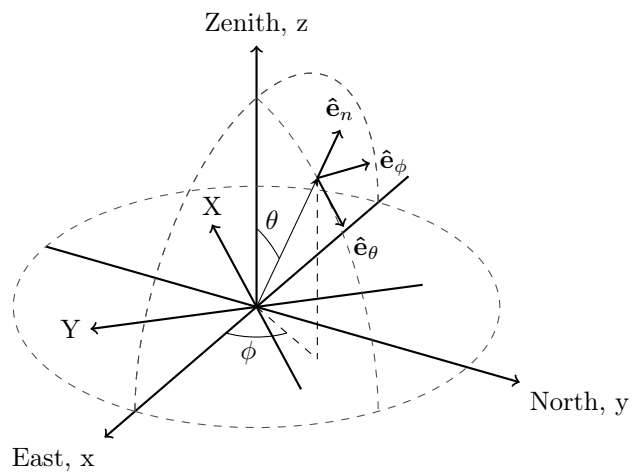
\begin{figure}
\centering
\tdplotsetmaincoords{60}{120}
\pgfmathsetmacro{\rvec}{0.8}
\pgfmathsetmacro{\thetavec}{30}
\pgfmathsetmacro{\phivec}{60}
\pgfmathsetmacro{\phivec}{60}
\begin{tikzpicture}[scale=4,tdplot_main_coords]
\coordinate (O) at (0,0,0);
\tdplotsetcoord{P}{\rvec}{\thetavec}{\phivec}
\draw[thick,->] (-1,0,0) -- (1,0,0) node[anchor=north east]{East, x};
\draw[thick,->] (0.4,0.4,0) -- (-0.4,-0.4,0) node[anchor=south]{X};
\draw[thick,->] (0,-0.8,0) -- (0,1,0) node[anchor=north west]{North, y};
\draw[thick,->] (-0.4,0.4,0) -- (0.4,-0.4,0) node[anchor=east]{Y};
\draw[thick,->] (0,0,0) -- (0,0,1) node[anchor=south]{Zenith, z};
\draw[-stealth] (O) -- (P) node[anchor=south west]{};
\draw[dashed] (O) -- (Pxy);
\draw[dashed] (P) -- (Pxy);
\tdplotdrawarc{(O)}{0.2}{0}{\phivec}{}{}
\node at (25:.3){$\phi$};
\tdplotsetthetaplanecoords{\phivec}
\tdplotdrawarc[tdplot_rotated_coords]{(0,0,0)}{0.3}{0}{\thetavec}{anchor = north}{}
\node[tdplot_rotated_coords] at (15:0.4){$\theta$};
\draw[dashed,tdplot_rotated_coords,color=dark-gray] (\rvec,0,0) arc (0:90:\rvec);
\draw[dashed,color=dark-gray] (\rvec,0,0) arc (0:360:\rvec);
\tdplotsetrotatedcoords{\phivec}{\thetavec}{0}
\tdplotsetrotatedcoordsorigin{(P)}
\draw[thick,tdplot_rotated_coords,->] (0,0,0) -- (.2,0,0) node[anchor=north west]{$\uvect{e}_{\theta}$};
\draw[thick,tdplot_rotated_coords,->] (0,0,0) -- (0,.2,0) node[anchor=west]{$\uvect{e}_{\phi}$};
\draw[thick,tdplot_rotated_coords,->] (0,0,0) -- (0,0,.3) node[anchor=south]{$\uvect{e}_{n}$};
\foreach \angle in {0,0}
{
\coordinate (P) at (0,0,\sintheta);
\tdplotsetthetaplanecoords{\angle}
\tdplotdrawarc[dashed,tdplot_rotated_coords,color=dark-gray]{(O)}{\rvec}{-90}{90}{}{}
}
\end{tikzpicture}
\caption[Geometry of coordinate systems]{On-sky polarization coordinate frame ($\uvect{e}_{\theta}$, $\uvect{e}_{\phi}$, $\uvect{e}_{n}$). Also depicted is the (north, east, zenith) coordinate frame of the simulations, where the unit vectors ($\uvect{e}_{x}$, $\uvect{e}_{y}$, $\uvect{e}_{z}$) correspond to the $x$, $y$ and $z$-axis, respectively. Furthermore the dipole antennas $X$ and $Y$ are shown.}
\label{fig:project}
\end{figure}

Any wave coming from a direction $\uvect{e}_{n}$ can be seen as a linear superposition of monochromatic plane waves, polarized in the $\uvect{e}_{\phi}$ and $\uvect{e}_{\theta}$ direction. Here $\phi$ and $\theta$ are the standard spherical coordinate angles with the $x$ and $z$ axis respectively, e.g.\
\begin{equation}
\vec{E}(t) = \sum_{\omega} \left(E_{\theta,\omega}\uvect{e}_{\theta} +  E_{\phi,\omega}\uvect{e}_{\phi}\right)e^{-i(k\vect{n}\cdot\vect{x}+\omega t)}.
\end{equation}
This geometry can be seen in Fig.~\ref{fig:project}.

These terms are related to the output voltage of the amplifier for each dipole, and for each frequency, via the Jones matrix $\matr{J}$ \citep{Jones:1941, Hamaker1996} of the antenna model
\begin{equation}
\begin{pmatrix}V_{X}\\ V_{Y}\end{pmatrix} = \begin{pmatrix}J_{X\theta} & J_{X\phi} \\ J_{Y\theta} & J_{Y_\phi}\end{pmatrix} \begin{pmatrix}E_{\theta}\\ E_{\phi}\end{pmatrix},
\end{equation}
where $J_{X\theta}$ is the complex response of the antenna and amplifier of the $X$-dipole to a wave purely polarized in the $\uvect{e}_{\theta}$ direction.

Therefore, in order to both correct for the antenna response and convert from output voltage to electric field strength in the on-sky frame (see Sect.~\ref{sec:coordinate_transformation}), each pair of Fourier components from the signal in the two instrumental polarizations ($X$,$Y$) is multiplied by the inverse Jones matrix, followed by an inverse Fourier transform back to the time domain.

\begin{figure}
	\centering
	\begin{subfigure}[b]{\figscale\textwidth}
		\includegraphics[width=\textwidth]{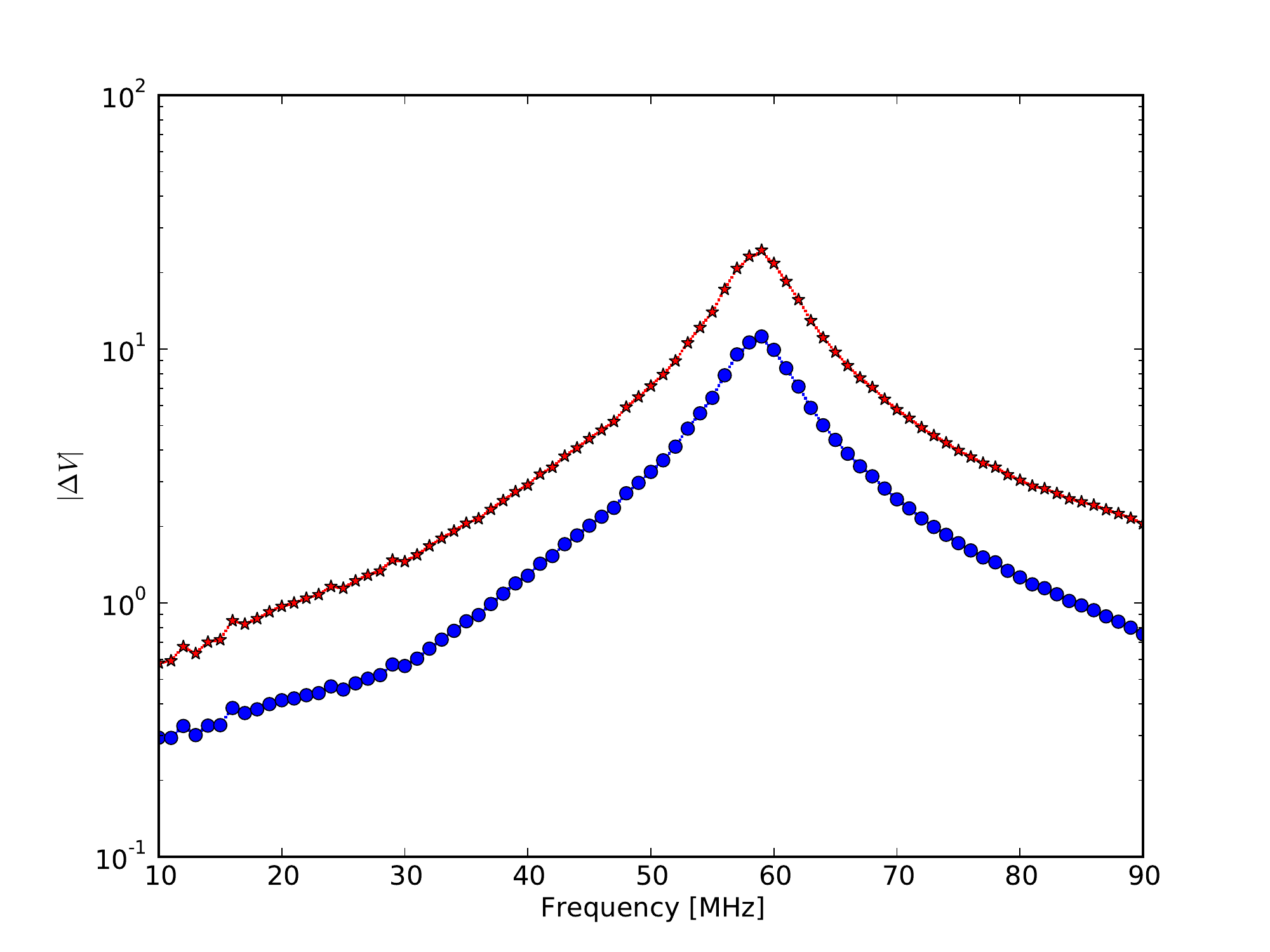}
		\label{fig:am_frequency_amplitude}
	\end{subfigure}
	\begin{subfigure}[b]{\figscale\textwidth}
		\includegraphics[width=\textwidth]{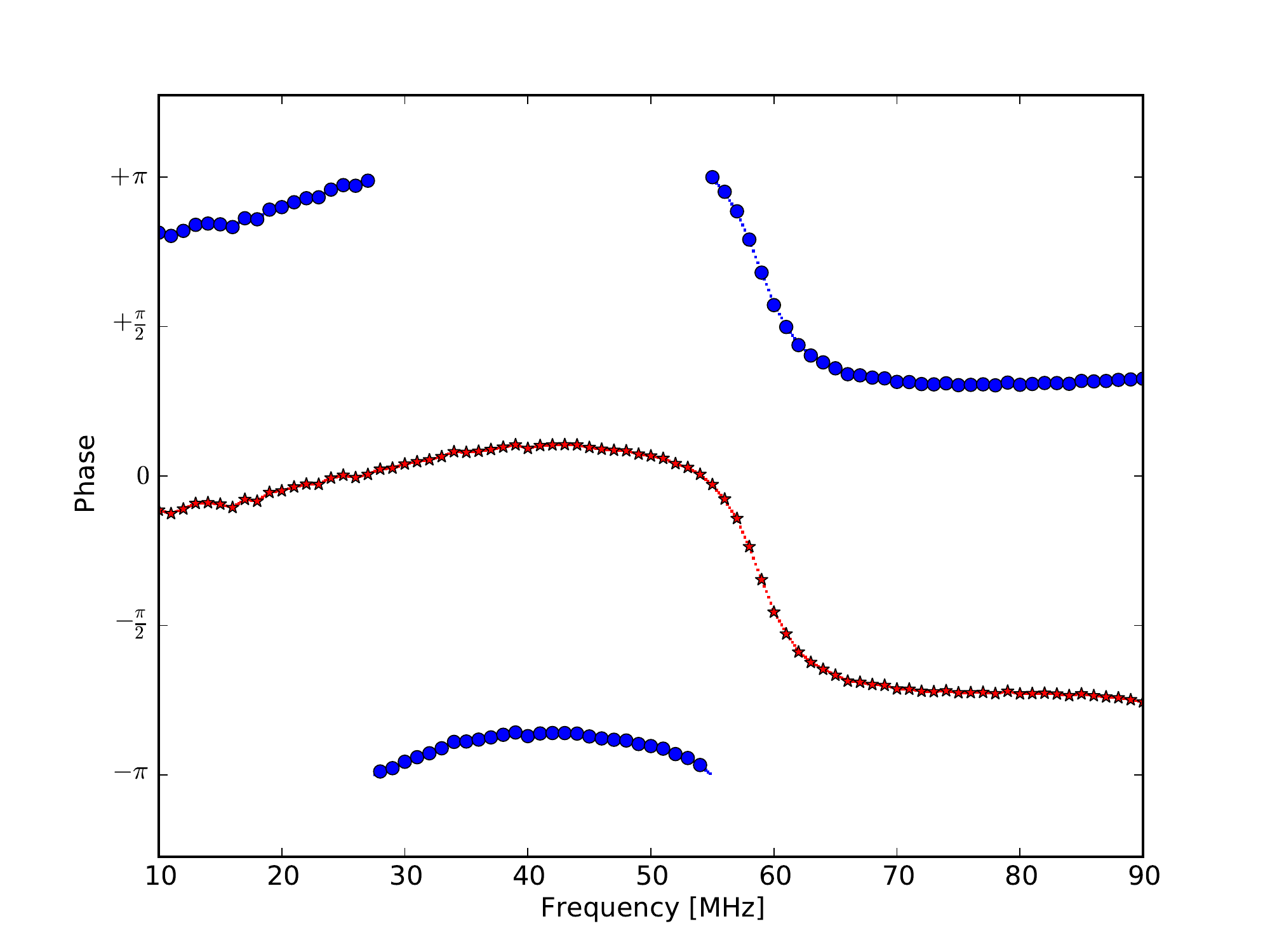}
	\label{fig:am_frequency_phase}
\end{subfigure}
\caption{Jones matrix components of the antenna model amplitudes (\emph{top}) and phases (\emph{bottom}) for a dipole receiving a wave polarized in the $\uvect{e}_{\theta}$ direction (circles) and a wave polarized in the $\uvect{e}_{\phi}$ direction (stars) for an arrival direction of $\phi=345^{\circ}, \theta=50^{\circ}$. Also plotted, as the dotted line, are the interpolated values.}
\label{fig:am_components}
\end{figure}

\begin{figure}
	\centering
	\includegraphics[width=\figscale\textwidth]{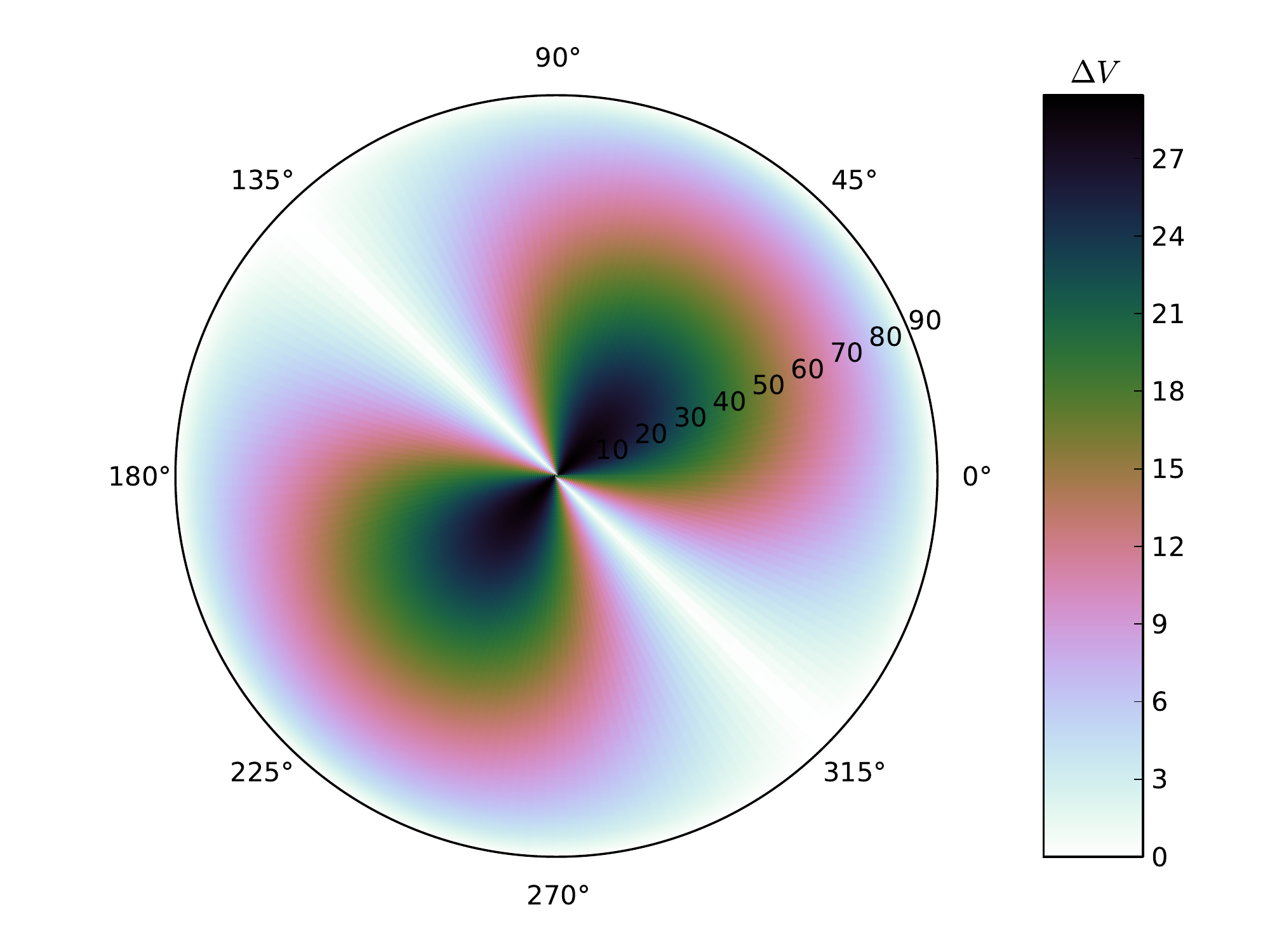}
	\caption[Directional dependence antenna pattern]{An example Jones matrix component describing the dipole response, at $\unit[60]{MHz}$, $|J_{X,\theta}|$ in the form of the output Voltage ($\Delta V$) as a function of direction for an incoming wave that is purely linearly polarized in the $\uvect{e}_{\theta}$ direction.}
\label{fig:am_amplitude_response_theta}
\end{figure}

The components of the Jones matrix of the antenna model are simulated on a grid with steps of $\unit[1]{MHz}$ in frequency, $5^{\circ}$ in $\theta$ and $10^{\circ}$ in $\phi$. In order to obtain the components at the frequency and direction of observation, trilinear interpolation is performed on the real and imaginary parts of the complex table when needed.
Examples of the response are depicted as a function of frequency in Fig.~\ref{fig:am_components} and as a function of direction in Fig.~\ref{fig:am_amplitude_response_theta}.

\subsubsection{Pulse detection}
Estimating the direction of the incoming air shower, see Sect.~\ref{sec:direction_fit}, can either be done using beamforming or through pulse timing. Beamforming was found to be very sensitive to the optimization algorithm used, essentially requiring a grid search to avoid getting stuck in a local minimum. This is computationally very expensive, moreover it only provides relative time differences between any two antennas rather than an absolute time needed for extraction of relevant physical parameters (see Sect.~\ref{sec:coordinate_transformation}).

In order to use pulse timing, individual pulses have to be identified. This can be done by using the cross-correlation method, where one looks for the maximum in the cross correlation of the signals between all antennas. This however has the same drawback as beamforming, as only relative timing is calculated. A method to retrieve the absolute pulse timing is through the use of the \emph{Hilbert envelope}, which is used in this pipeline. A detailed comparison of the methods is given in Sect.~\ref{sec:direction_fit}.

A sensible definition of the pulse arrival time is the measured arrival time of the maximum of the electric field strength. In practice, however, using directly $\max(|x(t)|^2)$ is highly dependent on the filter characteristics of the receiving system and the sampling used. Therefore, the arrival time is defined as the position of the maximum in the amplitude envelope of the analytic signal, also called the Hilbert envelope
\begin{equation}
A(t)=\sqrt{x^{2}(t) + \hat{x}^{2}(t)}.
\end{equation}
Where $\hat{x}(t)$ is the \emph{Hilbert transform}, or imaginary propagation, of the signal $x(t)$ defined by
\begin{equation}
\mathcal{F}(\hat{x}(t))(\omega) = -i\cdot \mathrm{sgn}(\omega)\cdot \mathcal{F}(x(t))(\omega)
\end{equation}
where $\mathcal{F}$ denotes the Fourier transform.

\begin{figure}
	\centering
\includegraphics[width=\figscale\textwidth]{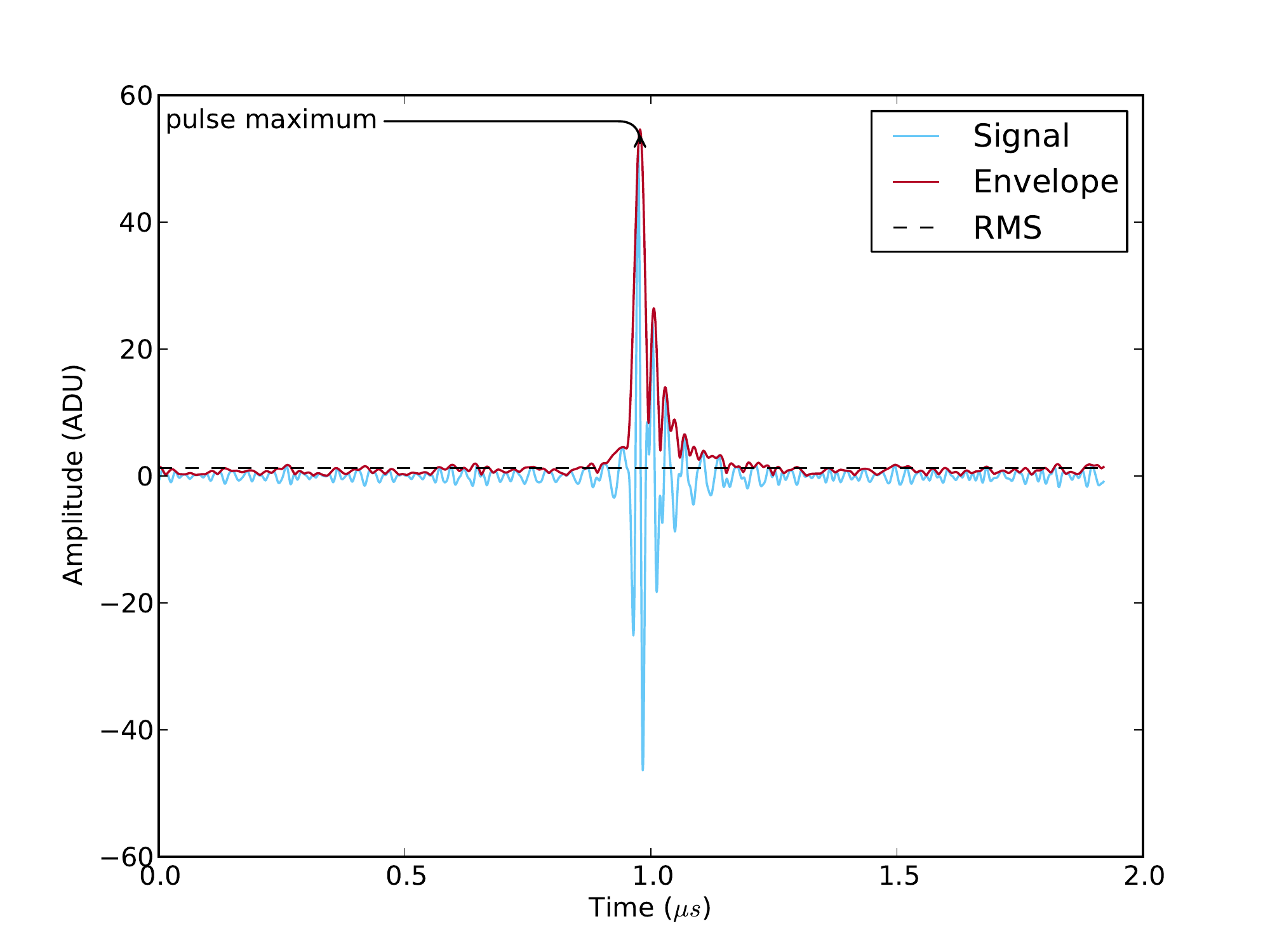}
	\caption[Pulse detection]{The solid light line shows the up-sampled signal. Overlaid is the Hilbert envelope and the RMS noise in black dashes. A pulse is accepted whenever the signal to noise ratio exceeds three.}
	\label{fig:example_pulse_detection}
\end{figure}

In order to find the pulse maximum with subsample precision, the signal is first up-sampled by a factor $16$, such that the maximum search will not be the limiting factor in the timing resolution. Subsequently, a simple maximum search is performed on the envelope. In addition, the signal-to-noise ratio is calculated where the signal is defined as the maximum and the noise as the root mean square of the envelope. An example can be seen in Fig.~\ref{fig:example_pulse_detection}.

This maximum search is performed on each of the on-sky polarizations $E_{\theta}(t)$ and $E_{\phi}(t)$ separately and any pulse with a signal to noise greater than three is marked as a possible cosmic-ray signal to be used for direction fitting. Because the pulse is expected to be intrinsically stronger in one of the two polarizations, depending on the angle between the shower axis and the geomagnetic field, the polarization with the highest average signal to noise (over all antennas) is first identified and only its maximum positions are used for the subsequent direction fit.

\subsubsection{Arrival direction fitting}
\label{sec:direction_fit}
\begin{figure}
	\centering
	\begin{subfigure}[b]{\figscale\textwidth}
	\includegraphics[width=\textwidth]{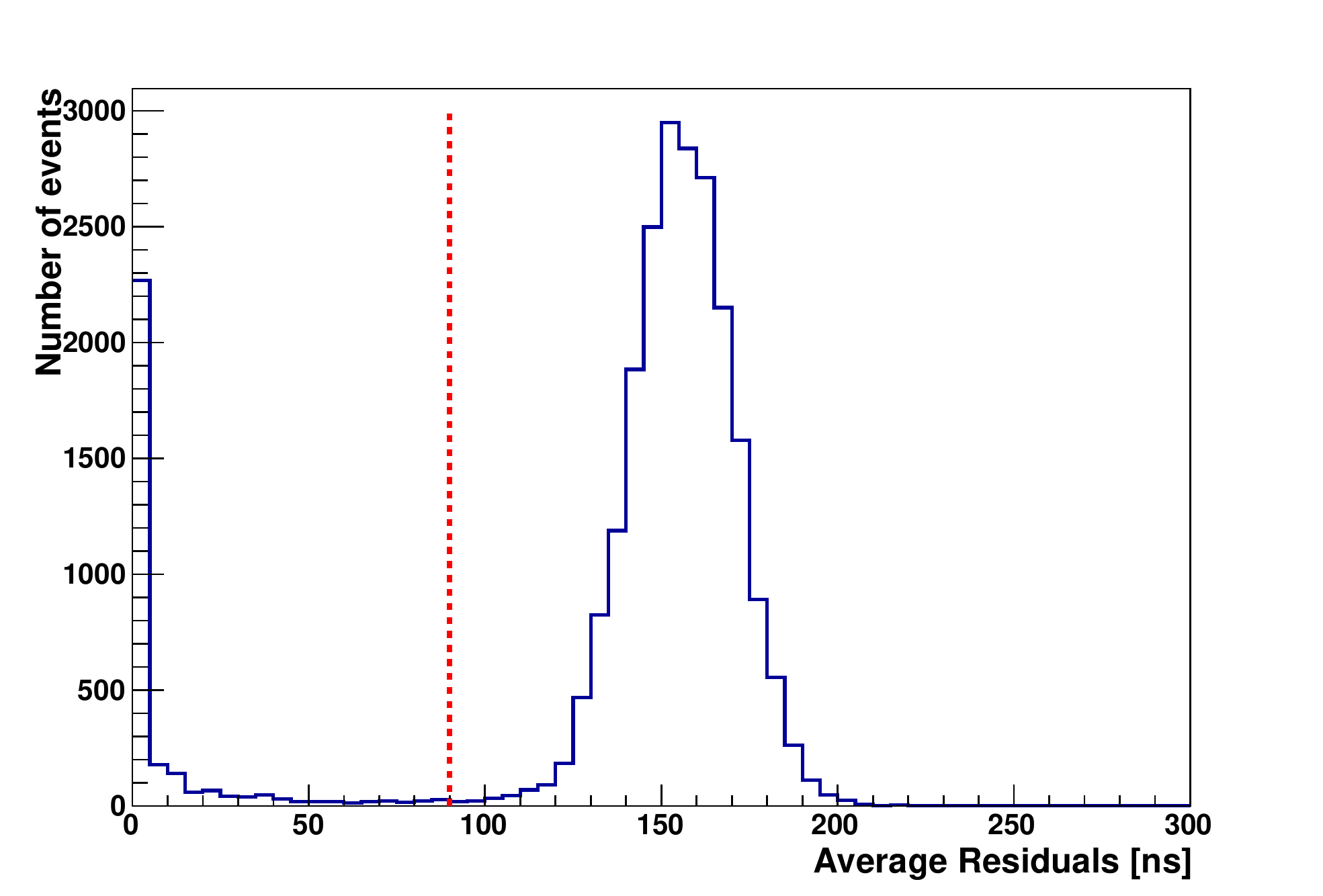}
	\end{subfigure}
\begin{subfigure}[b]{\figscale\textwidth}
	\centering
	\includegraphics[width=\textwidth]{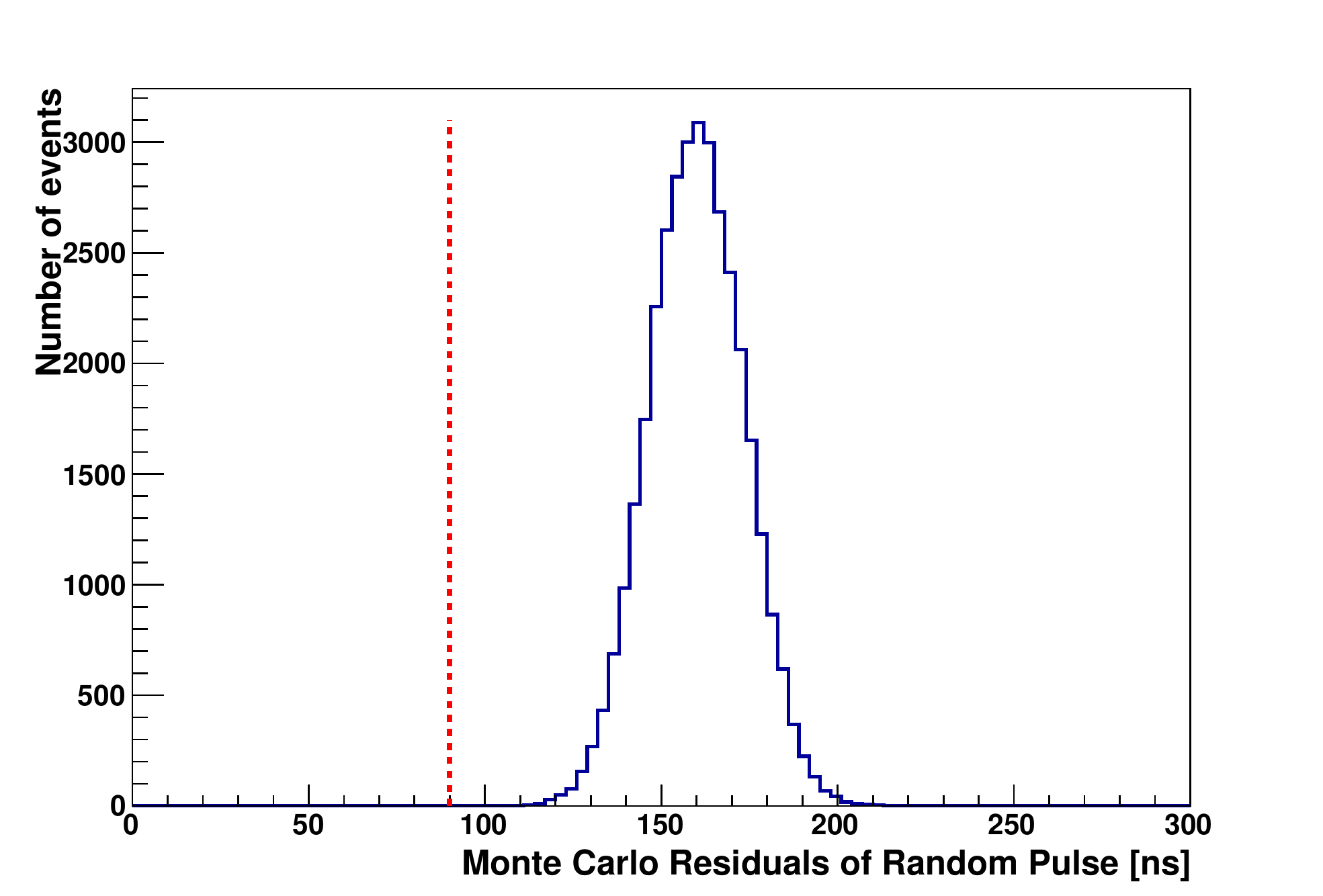}
\end{subfigure}
	\caption[Average residual delays MC]{Average residual delays derived from a plane-fit to data (\emph{top}) and from random samples in the search window with respect to a horizontal shower front (\emph{bottom}). The vertical line indicates the cut value derived from the simulated distribution, which is applied to the data.}
\label{fig:residual_distribution}
\end{figure}
As described above, every station is processed separately, meaning that the data do not provide a large lever arm for direction fitting. However, it also means that the actual shape of the shower front is an insignificant factor in the direction fitting. For a measurement with a single station, which has a maximum baseline of $\unit[80]{m}$, the shower front can be approximated by a plane wave. Thus, to determine the arrival direction of the cosmic ray a planar wavefront is fit to the arrival times of the pulses.

This method assumes that essentially all antennas are on a single plane, which certainly holds for all LOFAR stations as the ground was flattened during construction. Given a vector of arrival times $t$, and the vectors $x$ and $y$ for the coordinates of the antennas, the best fitting solution for a plane wave:
\begin{equation}
c t = A x + B y + C,
\end{equation}
can be found using a standard least squares approach. From $A$ and $B$ the Cartesian directions $\phi, \theta$ can be extracted as:
\begin{align}
A &=\sin(\theta) \sin(\phi),\\
B &=\sin(\theta) \cos(\phi).
\end{align}
The plane wave fit itself is done in several iterations. After a fit is performed the residual delays are investigated and those antennas that have residual delays larger than 3 times the standard deviation on the residual delays, are removed from the set and the data are refitted. The fit is terminated when there are less than four antennas left in the set or if no further antennas need to be removed. For this best direction all residual delays, including those of removed antennas, are calculated again and used for quality cuts later.

There are several quality criteria in the pipeline related to the plane wave fit. If the fit fails, a station is not considered further. In addition, a cut is made on the remaining average residual delays with respect to the expectation of the best fit. This cut can be derived from the distribution of all occurring plane wave residual delays, as shown in the top panel of Fig.~\ref{fig:residual_distribution}. The first peak with events of an average residual delay of less than $\unit[10]{ns}$ corresponds to excellent events, in which a clear cosmic-ray pulse can be identified in all antennas. The largest peak corresponds to all those events in which random noise fluctuations are identified as a pulse. This can be illustrated by a small Monte Carlo simulation. A random sample is picked from the range of the search window and its residual to the middle of the search window (corresponding to a vertical shower) is calculated. This results in the distribution in the bottom panel of Fig.~\ref{fig:residual_distribution}. The peak in the distribution obtained from data and the Monte Carlo distribution are centered around the same value and can therefore be identified with each other. Second order effects, being the directions of the air showers and non-infinite sampling, can influence the shape of the peak. The longer tail of the first peak (up to about $\unit[50]{ns}$) corresponds to events that have some antennas with correctly identified pulses and varying numbers of outliers, i.e.\ antennas where a random pulse is identified.

Therefore one can safely choose the value $\unit[90]{ns}$ as a first cut for good cosmic-ray events. Further cuts for higher quality events or stations can be applied in later analyses.
\begin{figure}
	\centering
	\includegraphics[width=\figscale\textwidth]{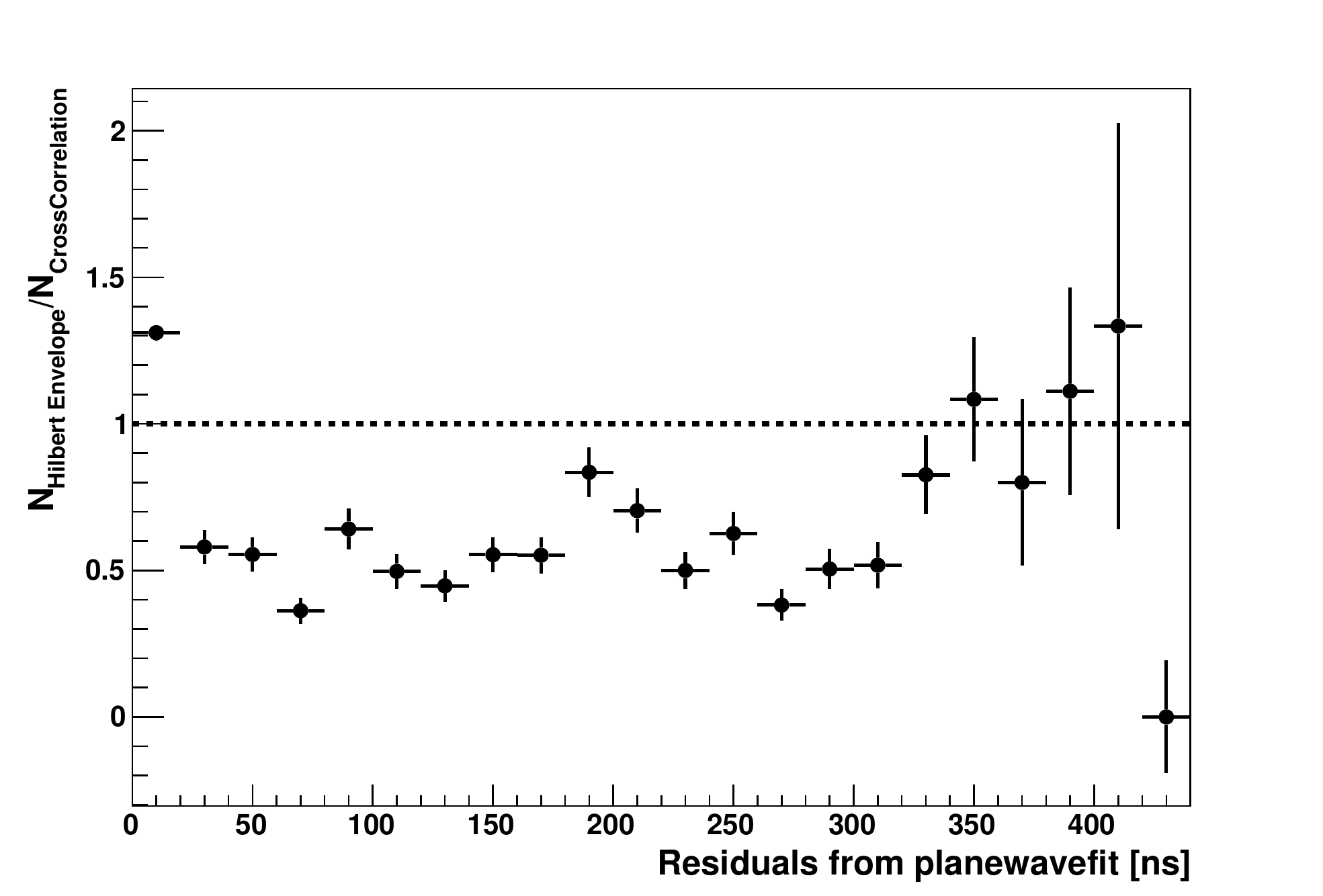}
	\caption[Residuals_Benchmark]{Difference in reconstruction between Hilbert Envelope and Cross Correlation. The different quality of the reconstruction is illustrated by plotting the fraction of the numbers of antennas $N$, identified by each method, with respect to the residual that was found in the plane wave reconstruction. For values above one the Hilbert Envelope identified more antennas, which is the case for the desired correctly identified signals, which can be found below $\unit[20]{ns}$.}
\label{fig:Res_bench1}
\end{figure}

The plane wave fit results now also allow for a justification of the choice of the Hilbert envelope as the method for pulse timing, as opposed to cross correlation. Figure \ref{fig:Res_bench1} shows the ratio of the number of antennas in which a pulse has been identified by either method with respect to the remaining residuals on a test-set of randomly chosen events that contain a cosmic-ray signal. The distribution clearly shows that the Hilbert envelope finds significantly more signals in the first bin, i.e.\ in the correct bin with small residuals. In general, cross correlating is expected to be better for pulses with lower signal-to-noise ratio. For pulses with a high signal-to-noise, however, the Hilbert transform performs more accurately. When using the Hilbert envelope, the position of the maximum is only determined by the recorded individual pulse, whereas the peak of the cross correlation is determined by the degree to which two signals correlate. This degree of correlation can be influenced by correlations in the noise (for instance residual RFI) or lacking similarity of the pulse shape between antennas, thereby making the cross-correlation less accurate for timing of pulses with a high signal-to-noise ratio.

\subsection{Coordinate Transformation}
\label{sec:coordinate_transformation}
After the antenna pattern unfolding cycle completes with a successful direction fit for a given station, the electric field components in on-sky polarizations $E_{\theta}(t)$, $E_{\phi}(t)$, and the shower arrival direction $\vect{n}$ are known. However, to compare measured data to air-shower simulations the three-dimensional electric field at ground level
\begin{equation}
\vect{E}(t) = E_{x}(t)\uvect{e}_{x} + E_{y}(t)\uvect{e}_{y} + E_{z}(t)\uvect{e}_{z}
\end{equation}
is needed, where $\uvect{e}_{x}$, $\uvect{e}_{y}$ and $\uvect{e}_{z}$ form the right handed coordinate system pointing east, north and up, respectively.
This geometry can also be seen in Fig.~\ref{fig:project}.

Assuming the signal has no electric field component in the propagation direction $-\uvect{e}_{n}$, this follows from a simple rotation $(E_{x}, E_{y}, E_{z})^{T} = \matr{R}\cdot (E_{\theta}, E_{\phi}, 0)^{T}$, with the rotation matrix
\begin{equation}
\matr{R}=\begin{pmatrix}
\cos\theta\cos\phi & -\sin\phi & \sin\theta\cos\phi\\
\cos\theta\sin\phi & \cos\phi & \sin\theta\sin\phi\\
-\sin\theta & 0 & \cos\theta
\end{pmatrix}.
\end{equation}

Note that this assumption is only an approximation, since the signal is measured in the near field of the shower and the source is moving. However, these are second order effects and the $E_{\theta}(t)$ and $E_{\phi}(t)$ components are expected to dominate over the $E_{n}(t)$ component \cite{RadioTheory}. Moreover, since LOFAR uses a dual polarization set-up it is not possible to extract the $E_{n}(t)$ component of a linearly polarized signal.

The pipeline concludes by storing pulse parameters for each antenna in the projected directions.

\subsection{Extracting Pulse Parameters}
\label{sec:extracting_pulse_parameters}
In addition to the shower arrival direction, obtained from pulse timing, two more parameters are extracted: for each antenna the peak amplitude and integrated power of the pulse are calculated.

Without multiplicative unit conversion factors, ignored for current lack of absolute calibration, the integrated pulse power is defined through the integration of the instantaneous Poynting vector and the electric field strength as:
\begin{equation}
P = \sum_{k} P_{k} \propto \sum_{k}\int_{\Delta t} |E_{k}(t)|^{2} dt,
\end{equation}
where $k=(x,y,z)$ are the polarization components of the electric field and $\Delta t$ is taken as a symmetric window around the pulse maximum.

This is calculated in discrete sampling $x_{i}$ as
\begin{equation}
P_{k} = \frac{1}{f}\left(\sum_{\mathrm{signal}} |x_{i}|^{2} - \frac{N_{\mathrm{signal}}}{N_{\mathrm{noise}}}\sum_{\mathrm{noise}} |x_{i}|^{2}\right),
\end{equation}
where $f = \unit[200]{MHz}$ is the sampling frequency and $N_{\mathrm{signal}}$ and $N_{\mathrm{noise}}$ are the number of samples in the signal and noise windows respectively. The noise window consists of the full $\unit[327680]{ns}$ block excluding the pulse window.
\begin{figure}
	\centering
	\includegraphics[width=\figscale\textwidth]{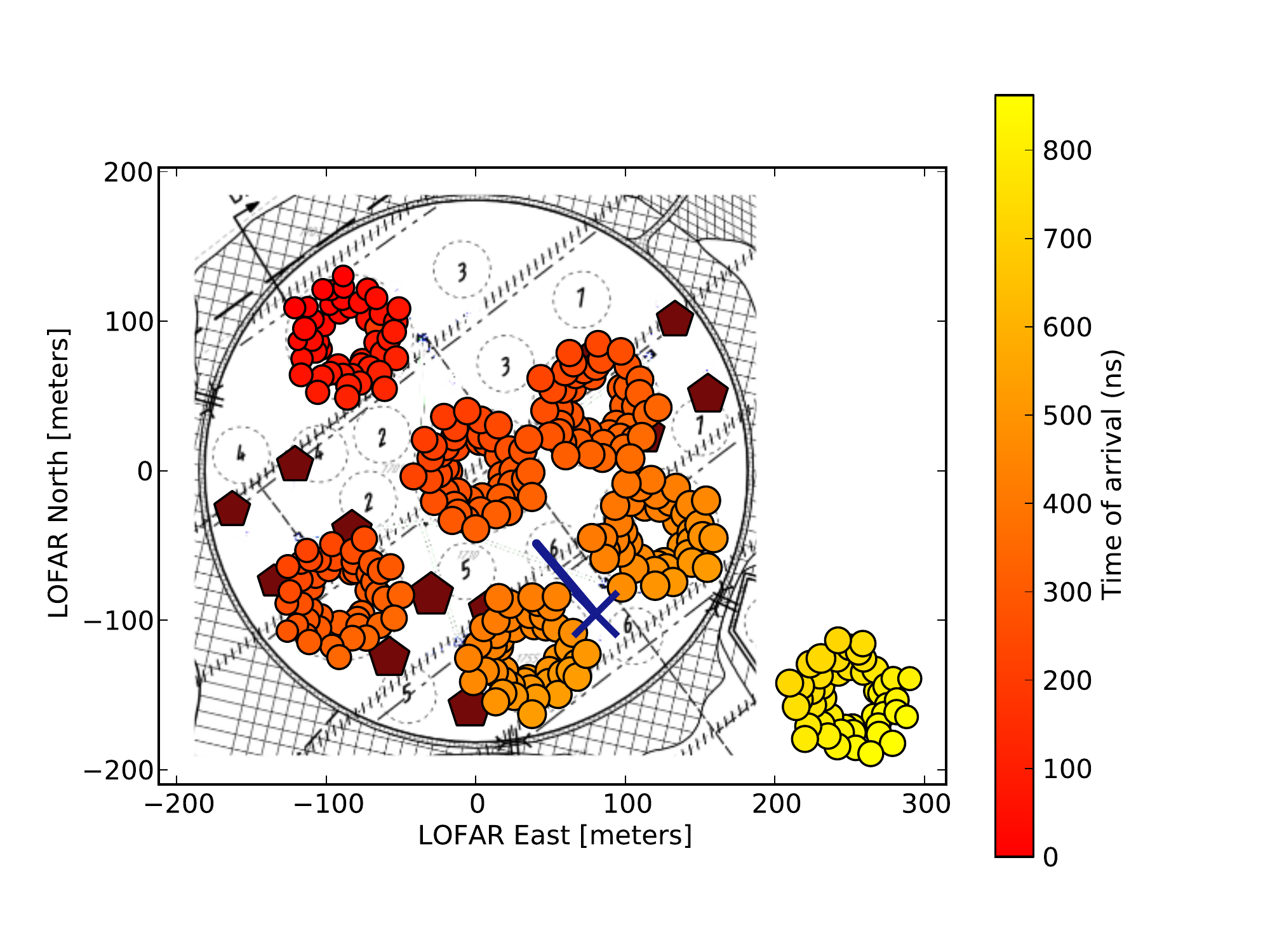}
	\caption[footprint]{Footprint of an air shower measured with LOFAR. The signal strength (peak amplitude of the radio signal) is encoded logarithmically in the size of the marker and the color shows the time of arrival. The pentagons represent the positions of the particle detectors, their size is proportional to the number of recorded particles. The reconstructed shower axis is indicated by the blue cross for the core position and the line for the projected arrival direction.}
	\label{fig:footprint}
\end{figure}

\begin{figure}
	\centering
	\includegraphics[width=\figscale\textwidth]{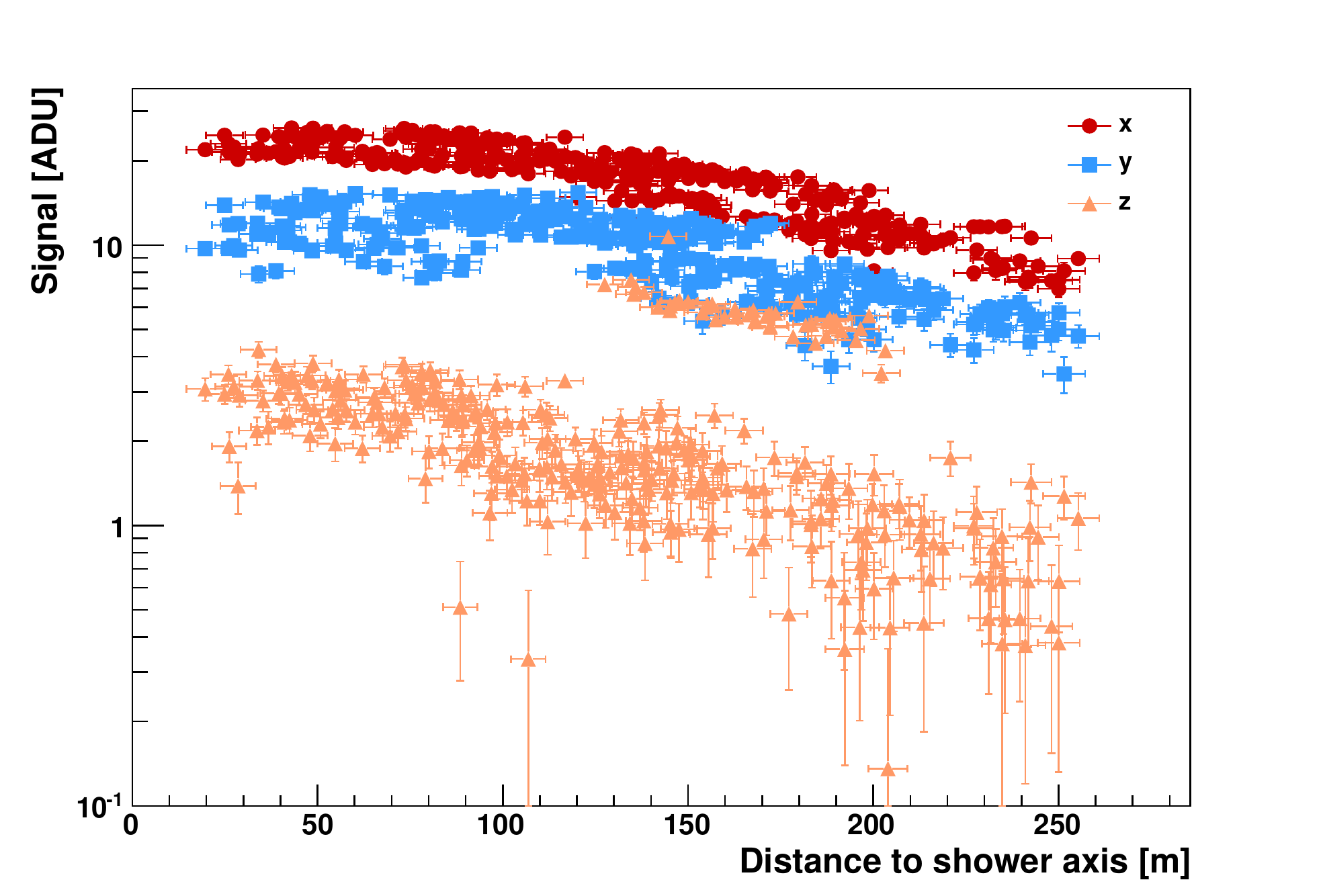}
	\caption[ldf]{Distribution of radio signals (peak amplitude in arbitrary units) with respect to the distance from the shower axis as reconstructed from the scintillator data. Shown are the three components of the reconstructed electric field. }
	\label{fig:ldf}
\end{figure}
\subsection{A measured air shower}
The result of the reconstruction pipeline is a full three-dimensional electric field vector per antenna position as a function of time. There are various ways in which this result can be visualized. The shower footprint, Fig.~\ref{fig:footprint}, shows the signal strength (peak amplitude of the radio signal) at the measured antenna locations as well as the time of arrival. Here, one can see that both the radio signal strength and the arrival times are consistent with the air-shower direction and core position as determined by the scintillator array. Both effects are distinctive properties of radio emission from air showers and are not produced by RFI.

Another common way to visualize the result is in the form of the lateral distribution, shown in Fig.~\ref{fig:ldf}. Here the radio signal strength, in all three polarization components, is shown as a function of projected distance to the shower axis. This projection retains the spatial distribution of the antennas (i.e.\ stations can be seen as groups), but azimuthal symmetry in the shower plane is assumed. This rather complicated looking distribution can be explained using detailed models of the radio emission, which also include non-rotational symmetrical effects. Further details of event by event characteristics will be reported in forthcoming publications.

\begin{figure}
	\centering
	\includegraphics[width=\figscale\textwidth]{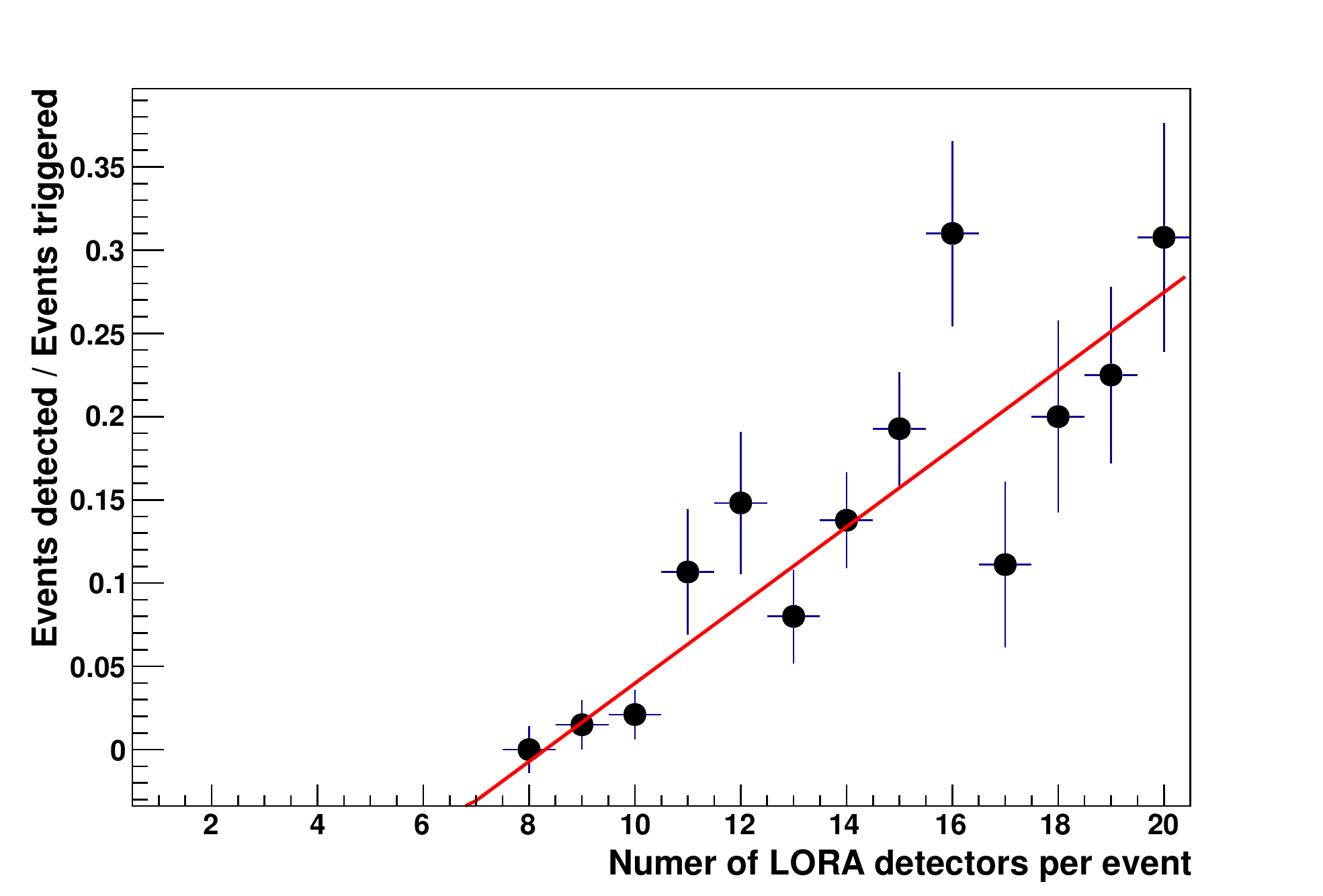}
	\caption[Trigger effectiveness]{Fraction of air showers with a detectable radio signal over the number of air showers triggered with a scintillator signal is plotted against the number of particle detectors above threshold in an event. The red straight line is a fit to the data.}
	\label{fig:triggeref}
\end{figure}

\section{Properties of reconstructed air showers}
\label{sec:dataset}
In order to verify the data quality and the method of reconstruction a short overview of the first data taken with LOFAR is given. The data set used here (June 2011 until April 2013) contains 3341 recorded triggers, of which 1597 pass the strict quality cut for a good data reconstruction of the particle measurement. Of all triggers, 405 events contain signals of cosmic rays as identified by the pipeline, with a threshold energy of $\unit[5\cdot10^{15}]{eV}$.

\subsection{Triggers from the array of particle detectors}
\label{sec:triggers}
On the reconstruction of air showers from the particle data quality cuts are applied. The reconstruction is considered reliable, when the reconstructed shower core is contained within the array, the shower is not too horizontal ($\theta < 50^{\circ}$) and the reconstructed Moli\`{e}re radius\footnote{Characteristic transverse size of an air shower.} falls in the range of $\unit[20-100]{m}$. After cuts, the lowest energy of a shower that triggered a read-out of the LOFAR buffers is $\unit[1.8\cdot10^{15}]{eV}$ and the highest is $\unit[1.9\cdot10^{18}]{eV}$. The LORA scintillator array becomes fully efficient above $\unit[2\cdot10^{16}]{eV}$.

All triggers sent by the scintillator array follow a nearly uniform distribution in azimuth and a $\sin(\theta)\cos(\theta)$-distribution in zenith angle as it is expected from the geometry for a horizontal array with flat detectors.

The number of events with a detectable radio signal increases with the number of triggered particle detectors, as can be seen in Fig.~\ref{fig:triggeref}, where the fraction of triggered events, with and without a detected radio signal, is plotted against the number of particle detectors per event. The fraction is clearly increasing with the number of triggered detectors, as shown by a fitted straight line. According to this fit, at a threshold of 13 detectors about 10\% of the events contain a cosmic-ray signal.

\begin{figure}
	\centering
	\includegraphics[width=\figscale\textwidth]{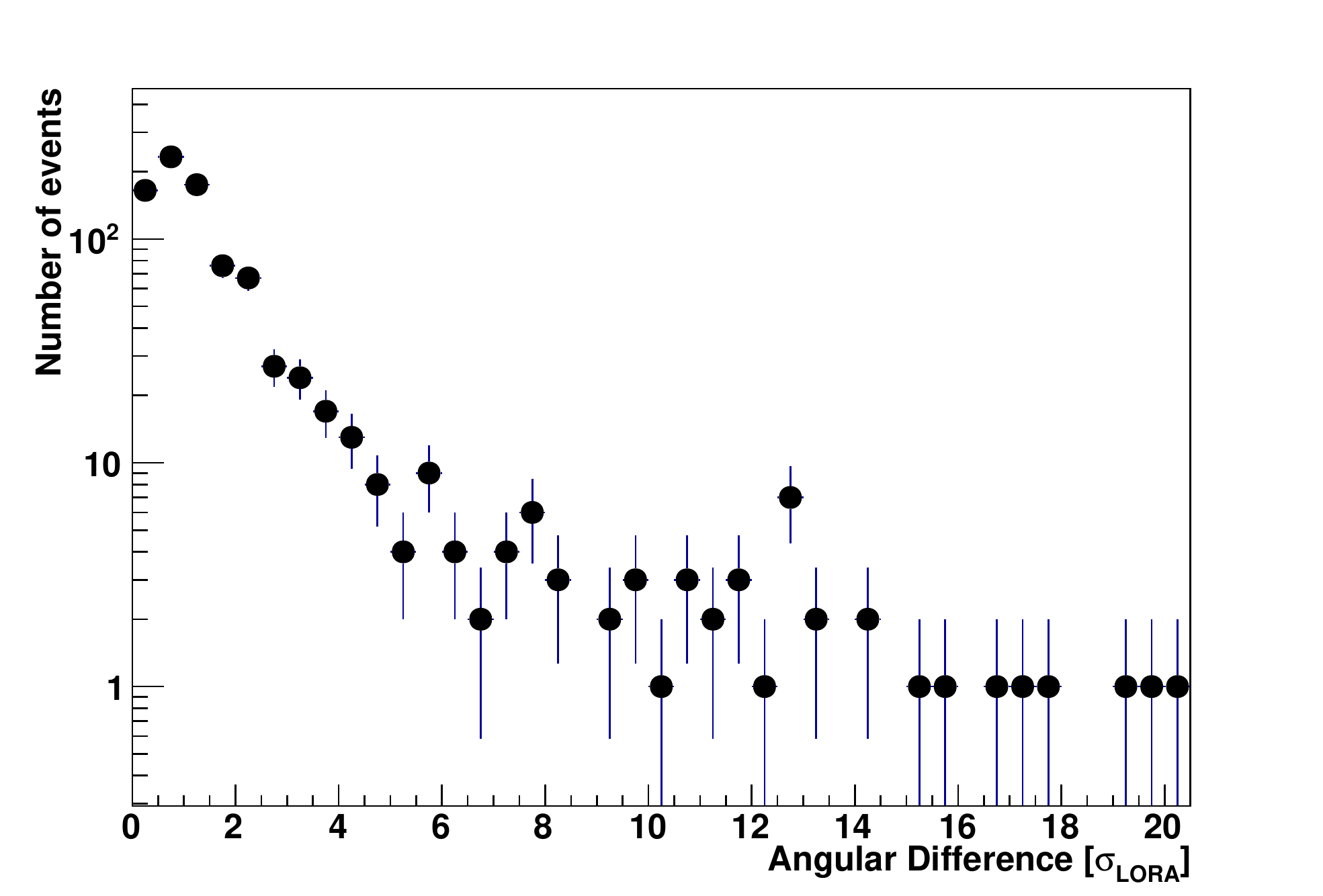}
	\caption[Azimuth distribution]{Angular difference between the shower axis reconstructed from the particle data and the direction estimate from the radio signal. To make the events comparable, the difference is scaled with the uncertainty of the individual reconstruction $\sigma_{\mathrm{LORA}}$.}
	\label{fig:sa}
\end{figure}

\subsection{Event rates and sensitivity}
For a first estimate all reconstructed triggers are considered valid events which show radio pulses coming from a direction that agrees to $10^{\circ}$ angular distance with the direction that was reconstructed from the arrival times measured with the particle detectors. This choice is based on the results shown in Fig.~\ref{fig:sa}. This figure shows the angular difference between the two reconstructed axes for all events. A steep fall-off in number of events with an increasing angular difference can be seen. Any event that deviates more than $\unit[10]{\sigma_{\mathrm{LORA}}}$ certainly lies outside the correct distribution. The shower axis is on average reconstructed with an uncertainty $\sigma_{\mathrm{LORA}} \sim 1^{\circ}$ from the data of the particle detectors. Thus, a quality cut of $10^{\circ}$ is chosen.

\begin{figure}
	\centering
	\includegraphics[width=\figscale\textwidth]{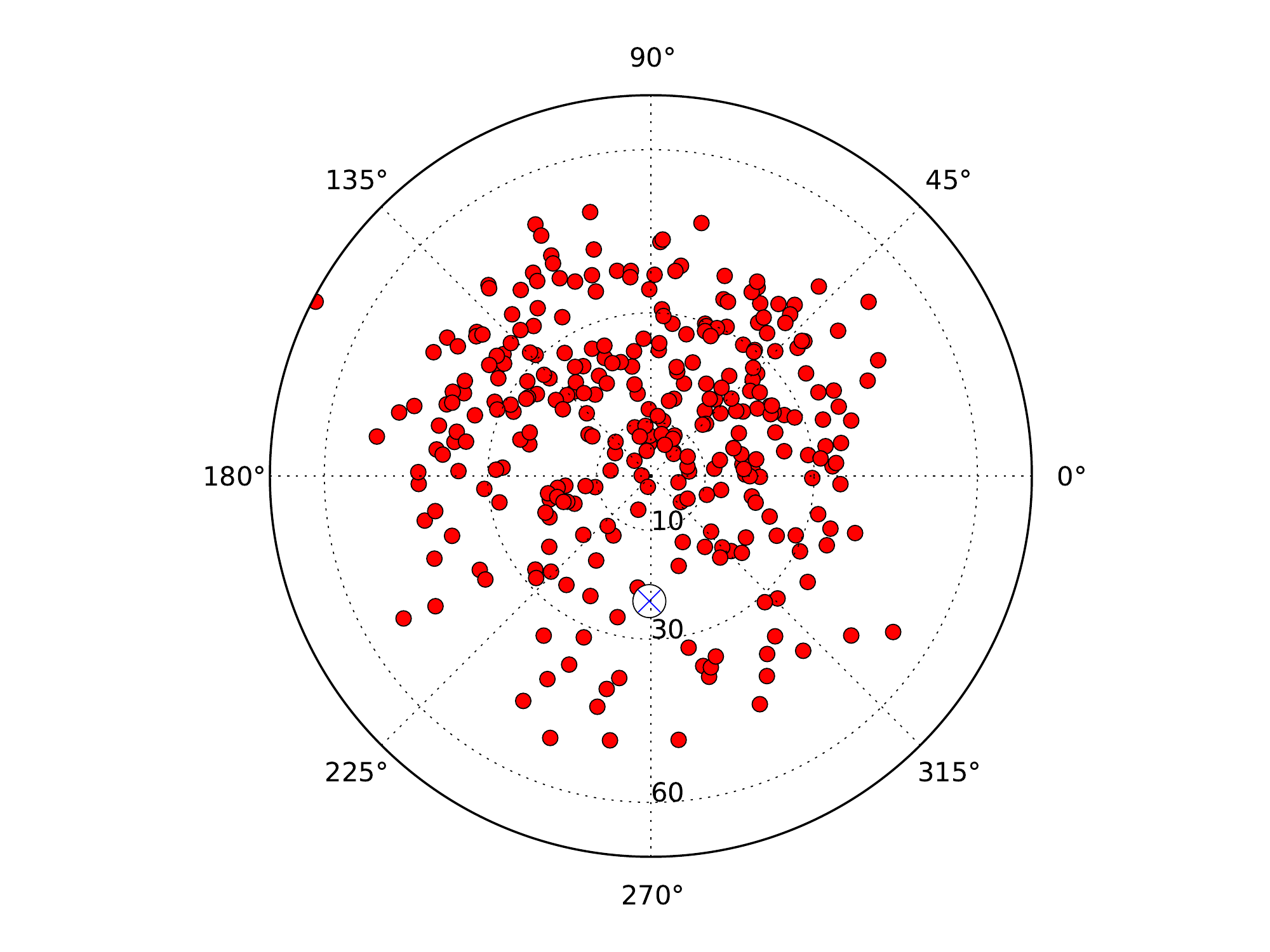}
	\caption[Sky plot]{Arrival directions of the cosmic-ray events detected with LOFAR from June 2011 until April 2013. East is $0^{\circ}$ and north corresponds to $90^{\circ}$. Also indicated (cross) is the direction of the magnetic field at LOFAR.}
	\label{fig:skyplot}
\end{figure}

Figure \ref{fig:skyplot} shows all 405 cosmic-ray events successfully detected with the LBAs as distributed on the local sky. Visible is a clear north-south asymmetry, where 276 events arrive from the northern hemisphere. This corresponds to a probability $p = 0.68 \pm 0.02$ for a detected event to arrive from the north. As the magnetic field at LOFAR is parallel to the north-south axis this is expected, if the main contribution to the signal is of geomagnetic origin \citep{Falcke2005,Codalema2009}.

\begin{figure}
	\centering
	\includegraphics[width=\figscale\textwidth]{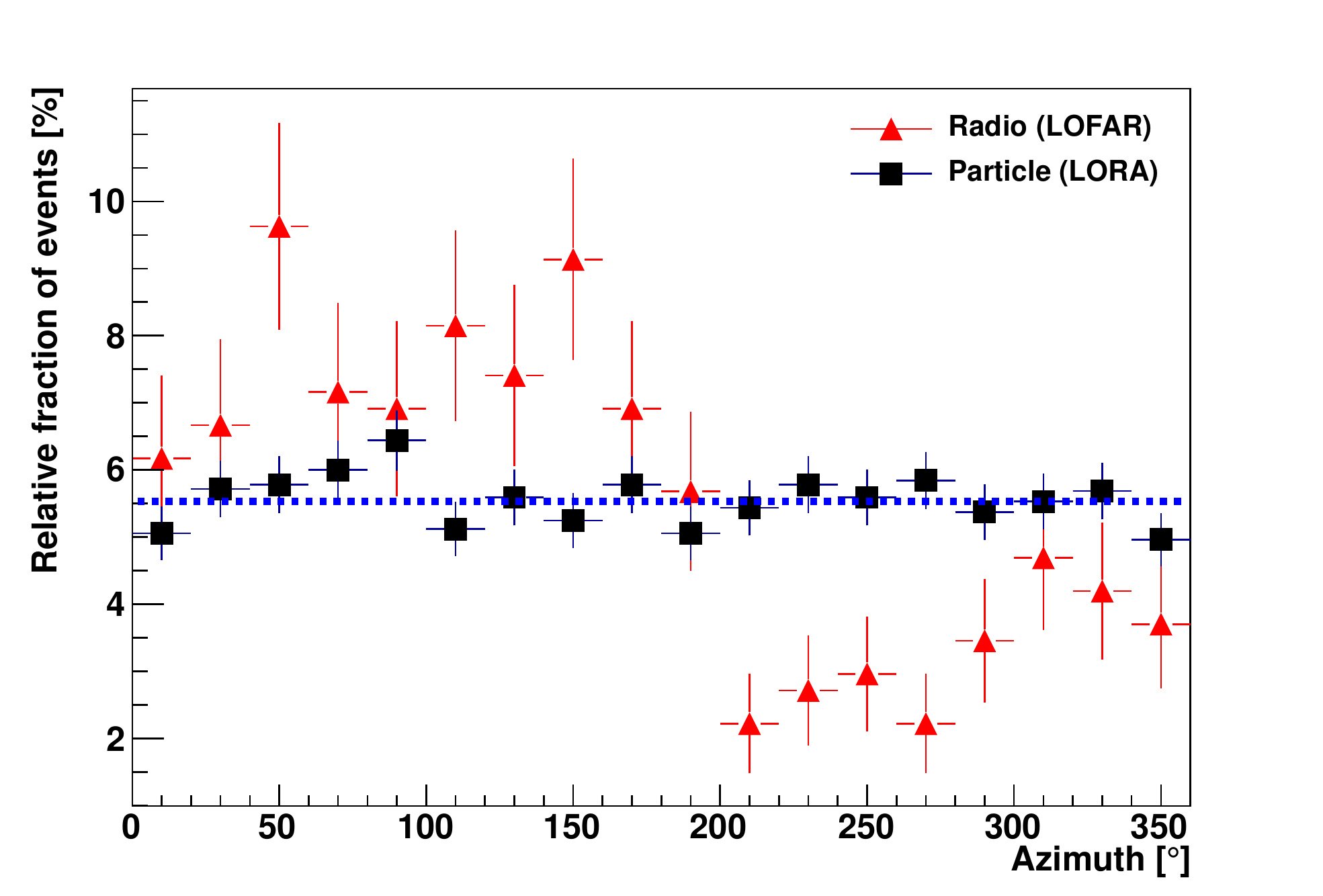}
	\caption[Azimuth distribution]{Binned distribution of the azimuth angles of all events measured with the particle detectors (black squares) and those in coincidence of particle detectors and radio antennas (red triangles). The best fit of a straight line to the particle data is also shown. The fit has a $\chi^2/\mathrm{nDoF}=0.9$.}
	\label{fig:azi}
\end{figure}

The effect is also illustrated in Fig.~\ref{fig:azi}, which shows the fraction of detected air showers as a function of azimuth angle for the events with radio signal, as well as for all LORA triggers sent. While the events registered with the LORA detectors are uniformly distributed in azimuth, the radio events show a clear deficit from the south. Due to the orientation of the LOFAR antennas and thereby the reduced sensitivity for purely east-west polarized signals, events arriving directly form the north are not necessarily preferred, as their signal is expected to be mainly polarized in the east-west direction \citep{RadioTheory}. The detection efficiency as a function of direction follows from a deconvolution of the expected emission strength with the antenna pattern and will not be discussed in detail here.

\begin{figure}
	\centering
	\includegraphics[width=\figscale\textwidth]{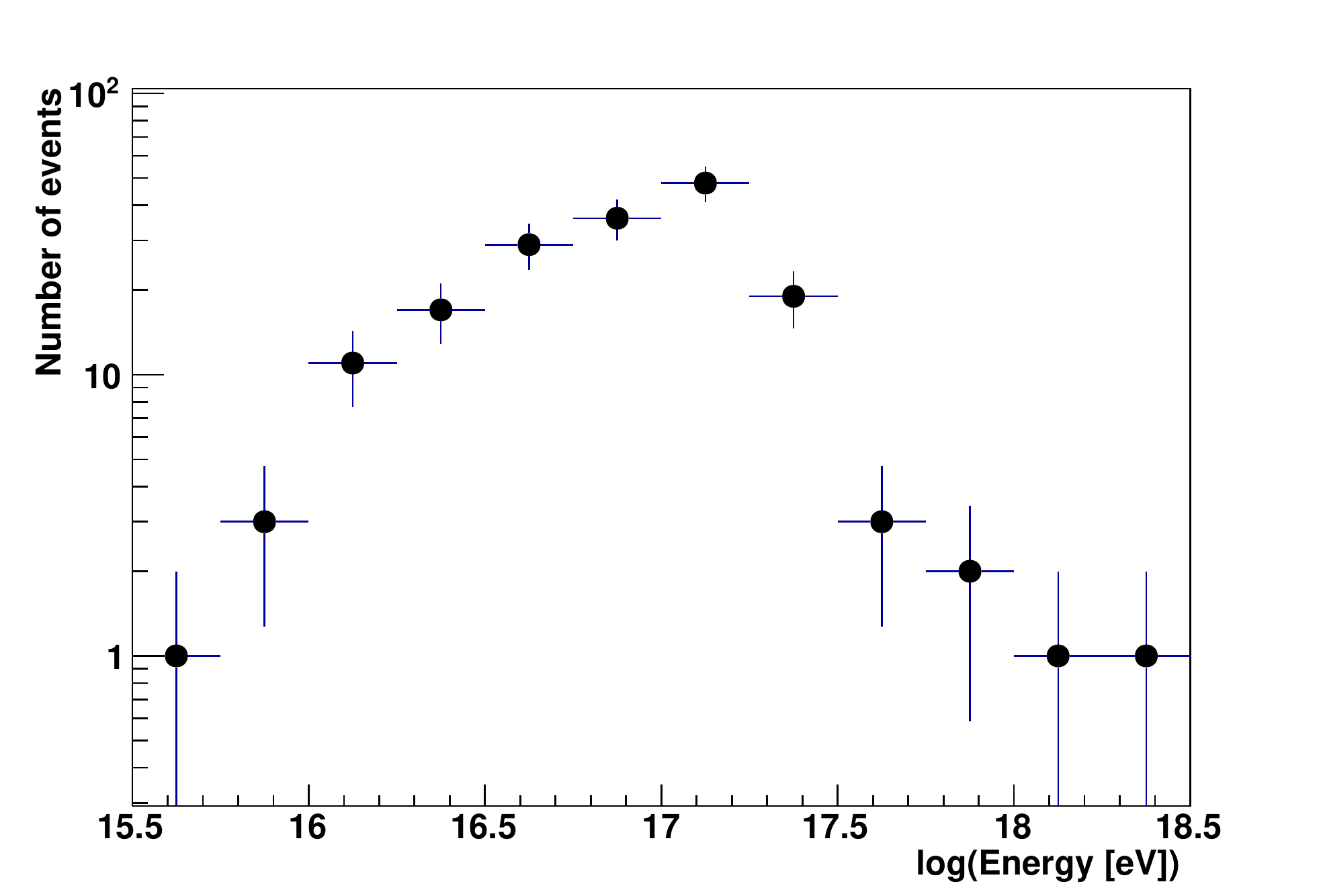}
	\caption[spectrum]{Distribution of the energies of the cosmic rays which had a measurable radio signal in the LOFAR data. The depicted energy is the one reconstructed from the corresponding particle data. The quality cuts, as described in Sect.~\ref{sec:triggers}, are applied.}
	\label{fig:spectrum}
\end{figure}

The energies of the air showers with a detectable radio signal are shown in Fig.~\ref{fig:spectrum}. The depicted energy is the one reconstructed from the corresponding particle data. This reconstruction has an overall systematic uncertainty of $27$\% and varying event by event uncertainties \citep{Thoudam2013}. One clearly sees that below $\unit[\sim10^{17}]{eV}$ the detection of air showers through their radio signal is not fully efficient, as the strength of the radio signal scales with the energy of the shower. Higher energies in this distribution are constrained by the steeply falling cosmic-ray energy spectrum and limited size of the detector array, which leads to limited event statistics at the highest energies. There are significant hints that showers of higher energies have been measured with LOFAR (especially when including the stations outside the Superterp), but these events are not well enough constrained by the data from the particle detectors in order to have a reference energy of the necessary accuracy. After a calibration of the energy of the radio measurements, those events will be used in a radio-stand-alone reconstruction.

\section{Conclusions}
\label{sec:conclusions}
At LOFAR cosmic-ray induced air showers are regularly measured with an array of particle detectors, LORA, and a large array of radio antennas. The cosmic-ray pipeline is routinely finding their distinctive radio signatures in the measurements and a full three-dimensional electric field vector is reconstructed for every antenna position.

A large dataset has been gathered with hundreds of identified cosmic-ray events in data from the LBAs. With up to a thousand antennas per events, these are the first highly detailed measurement of the radio signal of air showers. These measurements will be used for a detailed characterization of the shower shape and will be the benchmark data for comparison with models of radio emission in air showers.

\section*{Acknowledgements}
The LOFAR cosmic ray key science project very much acknowledges the scientific and technical support from ASTRON, especially in constructing the LORA particle detectors. We thank the KASCADE Collaboration for providing the scintillator detectors.
Furthermore, we acknowledge financial support from the Netherlands Research School for Astronomy (NOVA), the Samenwerkingsverband Noord-Nederland (SNN) and the Foundation for Fundamental Research on Matter (FOM) as well as support from the Netherlands Organization for Scientific Research (NWO), VENI grant 639-041-130. We acknowledge funding from an Advanced Grant of the European Research Council under the European Union's Seventh Framework Program (FP/2007-2013) / ERC Grant Agreement n. 227610.
LOFAR, the Low Frequency Array designed and constructed by ASTRON, has facilities in several countries, that are owned by various parties (each with their own funding sources), and that are collectively operated by the International LOFAR Telescope (ILT) foundation under a joint scientific policy.
The authors would like to thank both the internal and external referees for carefully reading the manuscript.
Chiara Ferrari and Giulia Macario acknowledge financial support by the
{\it "Agence Nationale de la Recherche"} through grant ANR-09-JCJC-0001-01.

\bibliographystyle{aa}
\bibliography{Detecting_cosmic_rays_with_the_LOFAR_radio_telescope}

\begin{thebibliography}{21}
\expandafter\ifx\csname natexlab\endcsname\relax\def\natexlab#1{#1}\fi

\bibitem[{{Alexov} {et~al.}(2012){Alexov}, {Schellart}, {ter Veen}, {van der
  Akker}, {B{\"a}hren}, {Greissmeier}, {Hessels}, {Mol}, {Renting}, {Swinbank},
  \& {Wise}}]{Alexov2013}
{Alexov}, A., {Schellart}, P., {ter Veen}, S., {et~al.} 2012, in Astronomical
  Society of the Pacific Conference Series, Vol. 461, Astronomical Data
  Analysis Software and Systems XXI, ed. P.~{Ballester}, D.~{Egret}, \&
  N.~P.~F. {Lorente}, 283

\bibitem[{Allan \& Jones(1966)}]{Allan1966}
Allan, H. \& Jones, J. 1966, Nature, 212, 129

\bibitem[{{Alvarez-Mu{\~n}iz} {et~al.}(2012){Alvarez-Mu{\~n}iz}, {Carvalho}, \&
  {Zas}}]{ZHAires}
{Alvarez-Mu{\~n}iz}, J., {Carvalho}, W.~R., \& {Zas}, E. 2012, Astroparticle
  Physics, 35, 325

\bibitem[{{Ardouin} {et~al.}(2009){Ardouin}, {Belletoile}, {Berat}, {Breton},
  {Charrier}, {Chauvin}, {Chendeb}, {Cordier}, {Dagoret-Campagne}, {Dallier},
  {Denis}, {Dumez-Viou}, {Fabrice}, {Gar{\c c}on}, {Garrido}, {Gautherot},
  {Gousset}, {Haddad}, {Koang}, {Lamblin}, {Lautridou}, {Lebrun}, {Lecacheux},
  {Lefeuvre}, {Martin}, {Meyer}, {Meyer}, {Meyer-Vernet}, {Monnier-Ragaigne},
  {Montanet}, {Payet}, {Plantier}, {Ravel}, {Revenu}, {Riviere}, {Saugrin},
  {Sourice}, {Stassi}, {Stutz}, \& {Valcares}}]{Codalema2009}
{Ardouin}, D., {Belletoile}, A., {Berat}, C., {et~al.} 2009, Astroparticle
  Physics, 31, 192

\bibitem[{{Ardouin} {et~al.}(2005){Ardouin}, {Belletoile}, {Charrier},
  {Dallier}, {Denis}, {Eschstruth}, {Gousset}, {Haddad}, {Lamblin},
  {Lautridou}, {Lecacheux}, {Monnier-Ragaigne}, {Rahmani}, \&
  {Ravel}}]{Codalema2005}
{Ardouin}, D., {Belletoile}, A., {Charrier}, D., {et~al.} 2005, International
  Journal of Modern Physics A, 20, 6869

\bibitem[{Falcke {et~al.}(2005)Falcke, Apel, Badea, B{\"a}hren, Bekk, \&
  Bercuci}]{Falcke2005}
Falcke, H., Apel, W.~D., Badea, A.~F., {et~al.} 2005, Nature, 435, 313

\bibitem[{{Falcke}(2008)}]{Falcke2008ICRC}
{Falcke}, H. e.~a. 2008, in ICRC Merida, Vol. Rapporteur, 30th International
  Cosmic Ray Conference

\bibitem[{{Hamaker} {et~al.}(1996){Hamaker}, {Bregman}, \&
  {Sault}}]{Hamaker1996}
{Hamaker}, J.~P., {Bregman}, J.~D., \& {Sault}, R.~J. 1996, \aaps, 117, 137

\bibitem[{Huege(2013)}]{RadioTheory}
Huege, T. 2013, ARENA 2012, AIP Conf. Proc. 1535, 121

\bibitem[{{Huege} {et~al.}(2012){Huege}, {Apel}, {Arteaga}, {Asch},
  {B{\"a}hren}, {Bekk}, {Bertaina}, {Biermann}, {Bl{\"u}mer}, {Bozdog},
  {Brancus}, {Buchholz}, {Buitink}, {Cantoni}, {Chiavassa}, {Daumiller}, {de
  Souza}, {Doll}, {Engel}, {Falcke}, {Finger}, {Fuhrmann}, {Gemmeke}, {Grupen},
  {Haungs}, {Heck}, {H{\"o}randel}, {Horneffer}, {Huber}, {Isar}, {Kampert},
  {Kang}, {Kr{\"o}mer}, {Kuijpers}, {Lafebre}, {Link}, {{\L}uczak}, {Ludwig},
  {Mathes}, {Melissas}, {Morello}, {Nehls}, {Oehlschl{\"a}ger}, {Palmieri},
  {Pierog}, {Rautenberg}, {Rebel}, {Roth}, {R{\"u}hle}, {Saftoiu}, {Schieler},
  {Schmidt}, {Schr{\"o}der}, {Sima}, {Toma}, {Trinchero}, {Weindl}, {Wochele},
  {Wommer}, {Zabierowski}, \& {Zensus}}]{Lopes2012}
{Huege}, T., {Apel}, W.~D., {Arteaga}, J.~C., {et~al.} 2012, Nuclear
  Instruments and Methods in Physics Research A, 662, 72

\bibitem[{{Huege} {et~al.}(2013){Huege}, {Ludwig}, \& {James}}]{CoREAS}
{Huege}, T., {Ludwig}, M., \& {James}, C.~W. 2013, ARENA 2012, AIP Conf. Proc.
  1535, 128

\bibitem[{Jelley {et~al.}(1965)Jelley, Fruin, Porter, Weekes, Smith, \&
  Proter}]{Jelley1965}
Jelley, J., Fruin, J., Porter, N., {et~al.} 1965, Nature, 205, 327

\bibitem[{{Jones}(1941)}]{Jones:1941}
{Jones}, R.~C. 1941, Journal of the Optical Society of America (1917-1983), 31,
  488

\bibitem[{Kolundzija(2011)}]{Kolundzija2011}
Kolundzija, B. 2011, in Proceedings of the 5th European Conference on Antennas
  and Propagation (EUCAP)

\bibitem[{{Marin} \& {Revenu}(2012)}]{Selfas}
{Marin}, V. \& {Revenu}, B. 2012, Astroparticle Physics, 35, 733

\bibitem[{{Nehls} {et~al.}(2008){Nehls}, {Hakenjos}, {Arts}, {Bl{\"u}mer},
  {Bozdog}, {van Cappellen}, {Falcke}, {Haungs}, {Horneffer}, {Huege}, {Isar},
  \& {Kr{\"o}mer}}]{Nehls2008}
{Nehls}, S., {Hakenjos}, A., {Arts}, M.~J., {et~al.} 2008, Nucl. Inst. Meth. A,
  589, 350

\bibitem[{{Offringa} {et~al.}(2013){Offringa}, {de Bruyn}, {Zaroubi}, {van
  Diepen}, {Martinez-Ruby}, {Labropoulos}, {Brentjens}, {Ciardi}, {Daiboo},
  {Harker}, {Jeli{\'c}}, {Kazemi}, {Koopmans}, {Mellema}, {Pandey}, {Pizzo},
  {Schaye}, {Vedantham}, {Veligatla}, {Wijnholds}, {Yatawatta}, {Zarka},
  {Alexov}, {Anderson}, {Asgekar}, {Avruch}, {Beck}, {Bell}, {Bell}, {Bentum},
  {Bernardi}, {Best}, {Birzan}, {Bonafede}, {Breitling}, {Broderick},
  {Br{\"u}ggen}, {Butcher}, {Conway}, {de Vos}, {Dettmar}, {Eisloeffel},
  {Falcke}, {Fender}, {Frieswijk}, {Gerbers}, {Griessmeier}, {Gunst},
  {Hassall}, {Heald}, {Hessels}, {Hoeft}, {Horneffer}, {Karastergiou},
  {Kondratiev}, {Koopman}, {Kuniyoshi}, {Kuper}, {Maat}, {Mann}, {McKean},
  {Meulman}, {Mevius}, {Mol}, {Nijboer}, {Noordam}, {Norden}, {Paas}, {Pandey},
  {Pizzo}, {Polatidis}, {Rafferty}, {Rawlings}, {Reich}, {R{\"o}ttgering},
  {Schoenmakers}, {Sluman}, {Smirnov}, {Sobey}, {Stappers}, {Steinmetz},
  {Swinbank}, {Tagger}, {Tang}, {Tasse}, {van Ardenne}, {van Cappellen}, {van
  Duin}, {van Haarlem}, {van Leeuwen}, {van Weeren}, {Vermeulen}, {Vocks},
  {Wijers}, {Wise}, \& {Wucknitz}}]{Offringa2013}
{Offringa}, A.~R., {de Bruyn}, A.~G., {Zaroubi}, S., {et~al.} 2013, \aap, 549,
  A11

\bibitem[{Thoudam {et~al.}(in prep.)Thoudam, Buitink, Corstanje, Enriquez,
  Falcke, Frieswijk, \& the LOFAR~CRKSP}]{Thoudam2013}
Thoudam, S., Buitink, S., Corstanje, A., {et~al.} in prep., Nuclear Instrument
  and Methods A

\bibitem[{{van Cappellen} {et~al.}(2007){van Cappellen}, Ruiter, \&
  Kant}]{LBADesign}
{van Cappellen}, W., Ruiter, M., \& Kant, G. 2007, LOFAR-ASTRON-ADD-009

\bibitem[{{van Haarlem} {et~al.}(2013){van Haarlem}, {Wise}, {Gunst}, {Heald},
  {McKean}, {Hessels}, {de Bruyn}, {Nijboer}, {Swinbank}, {Fallows},
  {Brentjens}, {Nelles}, {Beck}, {Falcke}, {Fender}, {H{\"o}randel},
  {Koopmans}, {Mann}, {Miley}, {R{\"o}ttgering}, {Stappers}, {Wijers},
  {Zaroubi}, {van den Akker}, {Alexov}, {Anderson}, {Anderson}, {van Ardenne},
  {Arts}, {Asgekar}, {Avruch}, {Batejat}, {B{\"a}hren}, {Bell}, {Bell}, {van
  Bemmel}, {Bennema}, {Bentum}, {Bernardi}, {Best}, {B{\^i}rzan}, {Bonafede},
  {Boonstra}, {Braun}, {Bregman}, {Breitling}, {van de Brink}, {Broderick},
  {Broekema}, {Brouw}, {Br{\"u}ggen}, {Butcher}, {van Cappellen}, {Ciardi},
  {Coenen}, {Conway}, {Coolen}, {Corstanje}, {Damstra}, {Davies}, {Deller},
  {Dettmar}, {van Diepen}, {Dijkstra}, {Donker}, {Doorduin}, {Dromer}, {Drost},
  {van Duin}, {Eisl{\"o}ffel}, {van Enst}, {Ferrari}, {Frieswijk}, {Gankema},
  {Garrett}, {de Gasperin}, {Gerbers}, {de Geus}, {Grie{\ss}meier}, {Grit},
  {Gruppen}, {Hamaker}, {Hassall}, {Hoeft}, {Holties}, {Horneffer}, {van der
  Horst}, {van Houwelingen}, {Huijgen}, {Iacobelli}, {Intema}, {Jackson},
  {Jelic}, {de Jong}, {Juette}, {Kant}, {Karastergiou}, {Koers}, {Kollen},
  {Kondratiev}, {Kooistra}, {Koopman}, {Koster}, {Kuniyoshi}, {Kramer},
  {Kuper}, {Lambropoulos}, {Law}, {van Leeuwen}, {Lemaitre}, {Loose}, {Maat},
  {Macario}, {Markoff}, {Masters}, {McFadden}, {McKay-Bukowski}, {Meijering},
  {Meulman}, {Mevius}, {Middelberg}, {Millenaar}, {Miller-Jones}, {Mohan},
  {Mol}, {Morawietz}, {Morganti}, {Mulcahy}, {Mulder}, {Munk}, {Nieuwenhuis},
  {van Nieuwpoort}, {Noordam}, {Norden}, {Noutsos}, {Offringa}, {Olofsson},
  {Omar}, {Orr{\'u}}, {Overeem}, {Paas}, {Pandey-Pommier}, {Pandey}, {Pizzo},
  {Polatidis}, {Rafferty}, {Rawlings}, {Reich}, {de Reijer}, {Reitsma},
  {Renting}, {Riemers}, {Rol}, {Romein}, {Roosjen}, {Ruiter}, {Scaife}, {van
  der Schaaf}, {Scheers}, {Schellart}, {Schoenmakers}, {Schoonderbeek},
  {Serylak}, {Shulevski}, {Sluman}, {Smirnov}, {Sobey}, {Spreeuw}, {Steinmetz},
  {Sterks}, {Stiepel}, {Stuurwold}, {Tagger}, {Tang}, {Tasse}, {Thomas},
  {Thoudam}, {Toribio}, {van der Tol}, {Usov}, {van Veelen}, {van der Veen},
  {ter Veen}, {Verbiest}, {Vermeulen}, {Vermaas}, {Vocks}, {Vogt}, {de Vos},
  {van der Wal}, {van Weeren}, {Weggemans}, {Weltevrede}, {White}, {Wijnholds},
  {Wilhelmsson}, {Wucknitz}, {Yatawatta}, {Zarka}, {Zensus}, \& {van
  Zwieten}}]{LOFAR}
{van Haarlem}, M.~P., {Wise}, M.~W., {Gunst}, A.~W., {et~al.} 2013, \aap, 556,
  A2

\bibitem[{{Werner} {et~al.}(2012){Werner}, {de Vries}, \& {Scholten}}]{EVA}
{Werner}, K., {de Vries}, K.~D., \& {Scholten}, O. 2012, Astroparticle Physics,
  37, 5

\end{thebibliography}

\end{document}